\documentclass{article}
\usepackage{graphicx} 
\usepackage{array} 
\usepackage[table]{xcolor}
\usepackage[a4paper, total={6in, 8in}]{geometry}

\title{Domestic frontier AI regulation, an IAEA for AI, an NPT for AI, and a US-led Allied Public-Private Partnership for AI: \\Four institutions for governing and developing frontier AI}
\author{Haydn Belfield
\\Leverhulme Centre for the Future of Intelligence 
\\University of Cambridge}
\date{September 2024}

\begin{document}

\maketitle

\begin{abstract} Compute governance can underpin international institutions for the governance of frontier AI. To demonstrate this I explore four institutions for governing and developing frontier AI. Next steps for compute-indexed domestic frontier AI regulation could include risk assessments and pre-approvals, data centre usage reports, and release gate regulation. Domestic regimes could be harmonized and monitored through an International AI Agency - an International Atomic Energy Agency (IAEA) for AI. This could be backed up by a Secure Chips Agreement - a Non-Proliferation Treaty (NPT) for AI. This would be a non-proliferation regime for advanced chips, building on the chip export controls - states that do not have an IAIA-certified frontier regulation regime would not be allowed to import advanced chips. Frontier training runs could be carried out by a megaproject between the USA and its allies - a US-led Allied Public-Private Partnership for frontier AI. As a project to develop advanced AI, this could have significant advantages over alternatives led by Big Tech or particular states: it could be more legitimate, secure, safe, non-adversarial, peaceful, and less prone to misuse. For each of these four scenarios, a key incentive for participation is access to the advanced AI chips that are necessary for frontier training runs and large-scale inference. Together, they can create a situation in which governments can be reassured that frontier AI is developed and deployed in a secure manner with misuse minimised and benefits widely shared. Building these institutions may take years or decades, but progress is incremental and evolutionary and the first steps have already been taken.

\end{abstract}

\pagebreak

\tableofcontents

\pagebreak
\section{Introduction}

Frontier AI regulation is now government policy in the USA, China, the EU and the UK. The USA's federal executive action and state bills, China’s Generative AI regulation, the EU AI Act, and the UK’s AI Safety Institute have profoundly changed the landscape in little over a year. Training runs that could create 'frontier' AI systems - those that are larger than any yet carried out (currently \(10^{26}\) FLOP) and that may pose threats to national and international security - should now be reported and tested with independent evaluations. This regime has been driven by governments’ profound national security concerns, especially that coming frontier AI systems may expand malicious actors’ ability to create cyber-weapons and even biological weapons. Evaluations will cover dual-use capabilities, societal impacts, system safety and security, and loss of control. These evaluations will be conducted by a range of bodies, primarily the government AI Safety Institutes (AISIs) in the International Network of AISIs - formed by the UK, USA, Singapore, Korea, Japan, Canada, Australia, France, Germany, Italy, and the EU AI Office. China has also committed in a Plenum document to establishing one, and there are several possibilities from the Network of AI Safety to the China Academy of Information and Communications Technology (CAICT) (Sheehan, 2023). This regime will be coordinated by AI Summits every six to twelve months and informed by the accompanying State of the Science reports led by the world’s most-cited AI scientist Prof. Yoshua Bengio. These evaluations will begin in earnest with OpenAI’s GPT-5, Google DeepMind’s Gemini 2, Anthropic’s Claude 4, and Meta’s Llama 4 - all of which are likely to be trained and evaluated in 2024/5. 

What might the next steps look like? How could these nascent regimes be made more robust and harmonised with one another, so that governments can have assurance that no insecure, unsafe, secret frontier training runs are carried out without their knowledge and oversight? How can participation be incentivised? More generally, how can governments promote the contribution of frontier AI models to peace, health and prosperity throughout the world while ensuring frontier AI is not developed or deployed in an unsafe, illegitimate or insecure manner? 

Both the international governance of AI and compute governance have recently received sustained scholarly and policy-maker attention (Ho et al., 2023; Sastry et al., 2024). This paper takes four AI governance scenarios that have been mentioned in the debate on AI governance and illustrates how compute governance can help implement them. This paper is not the first to propose these options, and does not aspire to be the final word on them. I explore them, flesh them out, and consider their positives and negatives.

I focus specifically on frontier AI: multimodal foundation models that are the largest yet in terms of training compute (as well as dataset size, parameter count, and cost) and that may pose several significant risks to national security. I note this is only a part of the AI field, and I do not intend to imply that other forms of AI or their governance are not also important. I make the key assumption that AI progress will continue the trend of the last decade, and arguably since the 1950s, and continue to be substantially driven by increases in computing power (compute). That is to say, ‘scaling laws’ will continue for the largest models - if trained on more compute and more data, performance will continue to reliably increase. 

This model of frontier AI development has certain advantages for governance - it relies on huge, visible, expensive infrastructure and a regular supply of state-of-the-art chips produced through concentrated supply chains - such that international frontier AI governance can focus on 5-10 of the biggest companies in the world. Indeed, this supply chain is more concentrated than for fissile material: each of the dual-use goods under the Nuclear Suppliers Group has 6-59 manufacturers worldwide whereas many steps in the AI supply chain only have 1-3 companies. This detectability, quantifiability and concentration – the sheer physicality, scarcity and cost of frontier AI training – makes the analogy with nuclear energy more appealing, and motivates my borrowing of models from the nuclear world for frontier AI. We have similar systems of domestic regulation, standard-setting, non-proliferation and public-private partnerships for other 'dual-use' industries that could be unsafe or misused: such as the pharmaceutical and biotechnological industries, the defence industry, and the chemical industry. For example, the Organisation for the Prohibition of Chemical Weapons carries out rigorous inspections.

If trends in compute usage and costs continue, frontier training runs will soon cost billions and then tens of billions of dollars. Rapid scaling of the largest frontier models can continue until at least 2030 (Sevilla et al., 2024). GPT-4 (2e25 FLOP) was trained on 10,000 times more computing power than GPT-2 (2e21 FLOP). By 2030, another 10,000x jump is possible to a 2e29 FLOP frontier training run. The data centre for this would use 10s of millions of  H100-equivalent GPUs and 5 gigawatts (GW) of energy - the equivalent of the largest nuclear power plants we have, the energy demand of a 1 million person city. This would require cumulative investments of hundreds of billions of dollars. This puts frontier AI training runs out of reach for almost any company in the world. No academic group, open source group or start-up is capable of raising that level of investment. Regulating frontier training runs therefore means regulating the very largest 'Big Tech' giants in the world, and those in ‘compute partnerships’ with them. 

Were this assumption not to hold - for example if scaling laws began to fail, electricity capacity was difficult to concentrate for these data centres, or there were significant algorithmic improvements or advances in distributed training - then these scenarios would be much less motivating.

I begin by exploring next steps for compute-indexed domestic AI regulation. These domestic regimes may soon require risk assessments and preapprovals for frontier training runs, usage reports and data centre audits, information security standards, and independent AI system evaluations conducted throughout training. Together, these can provide assurances to a government that companies are not conducting unsafe, insecure or secret frontier training runs within its territory. If a company does not abide by regulation, governments might not allow it to accumulate a large concentration of state-of-the-art chips. With these domestic regimes coming into place, I explore how they could be harmonized and monitored by an International AI Agency (IAIA) - an IAEA for AI. With a dual mandate, an IAIA should promote research, access and benefit-sharing while also developing safeguards and monitoring compliance. This could grow out of the cooperation of allies through the Network of AISIs and the wider AI Summits. This could potentially be backed up by a Secure Chips Agreement - an NPT for AI - building on the current chips export controls: only those countries that are compliant with IAIA safeguards would be allowed to import state-of-the-art-chips. Finally I explore how frontier training runs could be carried out by a US-led Allied public-private partnership for frontier AI. This could consolidate frontier training runs at facilities with extensive security controls. If cost trends continue, frontier training runs will be out of reach to all but the richest companies and states. A joint project could be more legitimate, secure, safe, non-adversarial, peaceful, and less prone to misuse than the two alternatives: Big Tech megaprojects or national megaprojects. Again a key incentive for participation is access to the advanced AI chips that are necessary for frontier training runs and large-scale inference.

The end point of these scenarios is ambitious. It would represent an international achievement similar to the international regime for nuclear energy - a secure, safe, legitimate, peaceful regime in which misuse is prevented and the benefits of a remarkable technology are broadly shared around the world. Establishing the international regime for nuclear energy took decades. The path to each of these scenarios is similarly long - yet it is also incremental, iterative and moderate. The structures already put in place will generate their own internal incentives and pressures for closing loopholes, harmonising standards and preventing free-riding.  The first steps towards an international regime for frontier AI have already been taken. Let us explore what could come next.

\pagebreak

\section{Compute-indexed domestic Frontier AI Regulation}

I begin by exploring the next steps for a domestic AI governance regime that leverages compute. How can a government ensure that frontier AI systems trained within its territory, or by companies headquartered in its territory, are safe and secure? 

The main emerging approach has been to index additional regulatory scrutiny to the amount of compute used to train an AI model (Anderljung et al., 2023; Shevlane et al., 2023; Heim and Koessler, 2024). This has been adopted explicitly by US Executive Order 14110, the EU AI Act, a draft for the Artificial Intelligence Law of the People’s Republic of China, and the work of the UK AISI. Specifically, the compute threshold of \(10^{26}\) FLOP training compute has emerged as a standard for additional scrutiny. This compute threshold is intended to rise over time so it only affects a small number of frontier systems. It would be more direct for a government to index regulatory scrutiny to specific AI capabilities and risks, but we often learn of AI capabilities only after the system is trained, or even deployed. This situation is untenable when serious harm can occur as a result of the AI system’s capabilities or the AI model can be stolen by malicious actors. So instead governments use compute thresholds as a proxy for capabilities and risks. The idea is simple: if a system is trained beyond a “risky threshold” of compute, then it is subject to increased proactive scrutiny and model evaluations, rather than solely focusing reactively on risks spotted during deployment. Such thresholds can also flag audits or other forms of risk assessment prior to starting the training run.

The end-goal is to reach a stable state where frontier AI training runs within these states are known, risk-assessed, safe and secure - where a government is reassured that no excessively risky frontier AI systems are being developed or deployed. By contrast, until recently governments only learnt about new cutting-edge models after a system was publicly announced, or when companies voluntarily elected to pre-notify. Risk assessment was based on evidence of harm already having occurred or speculation about specific domains where harm might occur; there was no systematic, standardized way to identify potentially high-risk capabilities ahead of time.

It is important that the level of risk be well understood by governments and that the level of oversight and regulation be proportionate to the level of risks, so that low-risk training runs are not unnecessarily caught and slowed. This is one of the key advantages of a compute threshold: training runs below the threshold would not face the same regulatory burden as those at or above it. The capacity to undergo regulatory scrutiny is likely to correlate with the overall resources (including compute) of an organization - compliance costs would be a small fraction of the overall cost of development. Indexing on compute has the advantage of a targeted and low regulatory burden for the AI industry: currently and in coming years, only a few companies are capable of frontier training runs, and only once or twice a year. The most capable systems take months to train and require a large fraction of a company’s computing power during that period.

Frontier AI regulation Bills have been proposed at a US Federal level, in US states like California, in the UK, and in China. The EU has passed the EU AI Act and is currently consulting on the Code of Practice for frontier AI systems.  These early steps in the USA, EU, China and the UK are still being developed and built. How might they evolve into a fully fledged regulatory regime? 

Requirements on AI companies can be grouped in four broad categories: to notify the government before a frontier training run, to inform the government about its physical security and cybersecurity, to report large data centres and large foreign cloud computing jobs, and to share the results of safety tests. Each of these four requirements can be built on and made more concrete, as I will discuss. There is other regulation that touches on frontier AI systems that I will not discuss in this section, such as copyright, liability and antitrust.

How can participation in such a frontier AI regulation regime be incentivised? Access to the latest generation of advanced AI chips could be conditional on participation in this frontier regulation scheme. State-of-the-art AI chips are in short supply. Governments could make companies' ability to accumulate the massive concentrations of compute required for such development conditional on compliance. Only frontier AI developers who submit to oversight would be authorized to train potentially risky frontier AI systems. If a company does not agree, it would not be allowed the privilege of accumulating the vast amount of state-of-the-art chips necessary for a frontier training run. Regulation (and the difficulty of accessing the required compute) would discourage unsafe, insecure or secret frontier runs.

\subsection{Next steps on frontier AI regulation}

A stylized path forward on frontier AI regulation, building on existing requirements, could proceed as follows.

\subsection{Phase 1: Risk assessments before frontier training runs}

First, companies are required to prenotify the government about planned frontier training runs. Over time, as AI systems become more capable and associated risks rise, prenotifying could eventually evolve into regulators requiring a risk assessment before frontier training runs. For example, a developer company might be required to submit a structured case that the training run is acceptably safe (Irving, 2024). Similar “safety cases” are required when building and operating nuclear power plants (IAEA, 2024). This proposal would likely involve evidence of expected capabilities and risks, details about information security, and plans for evaluations and testing. The company could receive feedback on this proposal from external, independent risk assessors or these mandatory risk assessments could be carried out by an ecosystem of third party auditors (CDEI, 2021). Risk assessment could leverage industry best-practices, and any standards or tools and benchmarks developed for these purposes, such as the sophisticated processes underway through CEN/CENELEC and NIST. 

In time, this could even take the form of preapproval from regulators for particular frontier training runs, similar to the preapproval regulators require for other experiments, such as certain high-risk biological experiments or rocket launches. In these other experiments, regulators have to be assured that risk assessment has been thorough and that safety and security risks are low. If regulators cannot be assured that the risks are sufficiently low enough, these training runs should not be run. A frontier training run is a significant, expensive and lengthy undertaking for a company preceded by a lot of planning and internal checks. Risk assessment is already a part of that planning - meeting best practice and sharing that risk assessment with a government would not represent a significant additional burden.

\subsection{Phase 2: Better cybersecurity, physical security and personnel security}

Second, companies are required to inform the government about the physical and cybersecurity protections taken to assure the integrity of the training process against sophisticated threats and to protect model weights from theft. Over time, as AI systems become more capable and associated risks rise, these protections could become more codified and stricter. For example, RAND defines five security levels: SL1 to SL5 (Nevo et al., 2024). No companies are currently meeting SL3, so they cannot “likely thwart cybercrime syndicates or insider threats”, let alone “operations by leading cyber-capable institutions” like states. In time, companies may need to seek advice and support from their state’s intelligence and security agencies - especially at the highest end of SL4 and SL5 to be secure against a state actor. 

\subsection{Phase 3: Usage reports and data centre audits}

Third, US companies are required to notify the US Government about large US data centres, frontier training runs, and large foreign cloud computing jobs conducted on data centres owned by US companies. This is intended to prevent the possibility of an undetected and unauthorised frontier training run. Over time, to provide further evidence that this is not occurring, auditors of the largest data centres could receive ongoing, detailed, anonymized usage reports. This should be done in cost-effective ways that preserve privacy and efficiency. For example, one proposal is to send cryptographic challenges to random chips - if they do not send the response back in time, then that chip may be being used for an unnotified training run (Shavit, 2023). The auditor receives no information about the content of the computation, just whether the chip is being used or not. Frontier training runs that fail to comply with set standards (such as prenotification) should be halted. Depending on their culpability, the labs and data centres responsible for them may incur sanctions.

In time, this could even involve inspections by regulators. For example, auditors may want to verify that no AI chips have been tampered with - including no tampering with potential on-chip features such as secure enclaves or location tracking. The inspections should tally with the usage reports. This will also be important in time to demonstrating compliance with the export controls - that there has been no diversion or smuggling of state-of-the-art chips.

\subsection{Phase 4: Release gate regulation}

Fourth, companies are required to share the results of safety tests carried out by the companies. Over time, these tests and evaluations will also be carried out by independent third parties such as the AISIs. This will require structured access to the model, and perhaps the weights themselves. The technical specifics of the evaluations and tests are currently being developed by the various AISIs, and by the NIST and CEN/CENELEC standardization processes. 

In time, regulators could decide whether to authorize deployment of potentially high-risk frontier models based on independent evaluations and tests. If these tests do not demonstrate that the model (and accompanying deployment mitigations) would be acceptably safe and secure, regulators could require companies to run more tests, to implement more deployment mitigations, or to do more post-training and fine-tuning with different datasets, as proposed in the EU AI Act. Moreover, evaluations could be run \textit{during }as well as \textit{after} frontier training runs to ensure consistent application of safety standards and identify possible threats at an earlier stage (METR, 2023; Shevlane et al., 2023).

\begin{figure}
    \centering
    \includegraphics[width=1\linewidth]{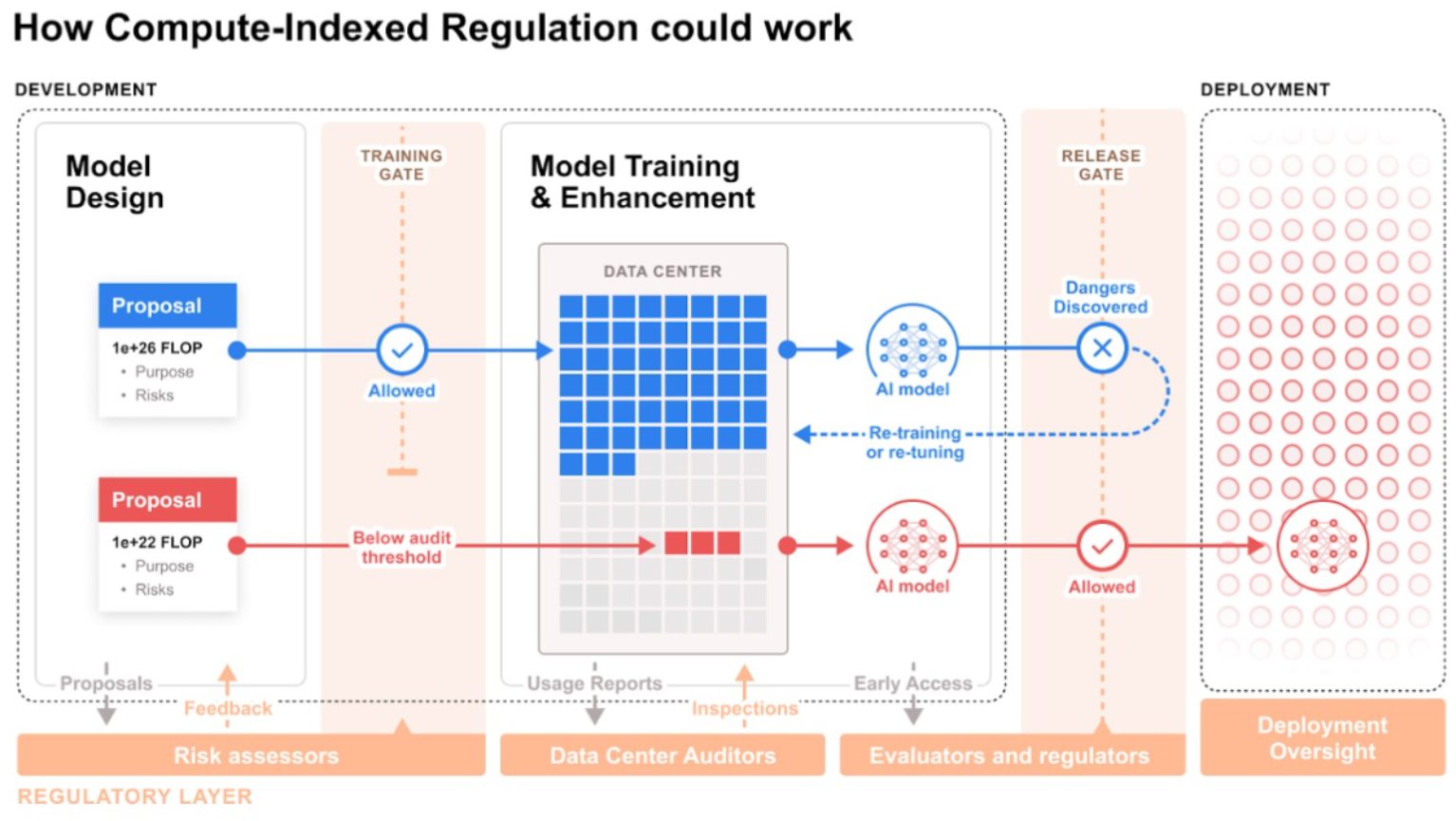}
    \caption{\textbf{How Compute-Indexed Regulation Could Work} A frontier training run over a certain compute threshold (blue) is proposed to an independent risk assessor before it is run on an physically secure, cyber secure and audited data centre. A smaller training run below the compute threshold (red) does not need to be prenotified. Both systems are tested before they are released. There is then deployment oversight for unexpected uses, capabilities and risks.}
    \label{fig:1}
\end{figure}

\pagebreak

\section{International AI Agency (IAIA) - an ‘IAEA for frontier AI’}

I now explore next steps for harmonising domestic frontier AI governance regimes. How can governments promote the contribution of frontier AI models to peace, health and prosperity throughout the world while ensuring frontier AI is not developed or deployed in an unsafe or insecure manner? 

As discussed in the previous section, there are nascent domestic frontier AI regulation regimes in the USA, UK, EU and China. These domestic frontier AI regimes are developing national safeguards. There are likely to be strong incentives to harmonise the national safeguards of these different domestic frontier AI regimes. Governments will likely want to ensure they are not falling behind the state of the science or not meeting best practice, and that other states are doing the same - that other states are not 'free-riding' or undercutting them. The AISIs will likely want to coordinate their approach to conducting risk assessments, evaluations and uplift studies. We already see the beginning of this harmonisation and coordination. Several regimes have adopted compute-indexed regulation, and the compute thresholds have all been set at a similar level, around \(10^{26}\) FLOP. Australia, Canada, France, Germany, Italy, Japan, Korea, Singapore, the UK, the USA and the EU have agreed to form an International Network of AISIs through the Seoul Declaration (DSIT, 2024b, 2024a). Moreover, several AISIs have signed Memorandums of Understanding (DSIT, 2024d).

Multinational frontier AI companies will also likely want some degree of standardisation between these regimes to reduce regulatory burden. For example, companies will not want to be required to provide slightly different information in a dozen different ways, and will want to avoid the expensive requirement of having to carry out several frontier training runs to comply with different regimes. 

Such harmonisation could be aided by an International AI Agency (IAIA) - an ‘IAEA for AI’. The International Atomic Energy Agency (IAEA) carries out three main tasks: developing nuclear safeguard standards; conducting nuclear monitoring and inspections; and promoting nuclear research, access and benefit-sharing. Similarly, the IAIA could carry out three main tasks: developing frontier AI safeguard standards; conducting frontier AI monitoring and inspections; and promoting frontier AI research, access and benefit-sharing.  This third task is a core part of the IAEA’s notable ‘dual mandate’, as defined in Article II of the IAEA Statute:

\begin{quote}“The Agency shall seek to accelerate and enlarge the contribution of atomic energy to peace, health and prosperity throughout the world. It shall ensure, so far as it is able, that assistance provided by it or at its request or under its supervision or control is not used in such a way as to further any military purpose”.    
\end{quote}

In the same way, the IAIA should have a dual mandate - to accelerate and enlarge the contribution of frontier AI models to peace, health and prosperity throughout the world while ensuring, so far as it is able, that frontier AI is not developed or deployed in an unsafe or insecure manner. The assurances provided by the standards and monitoring conducted by an IAIA would help enable safe and secure frontier training runs. The IAIA would not seek to \textit{prevent} frontier AI development but rather to \textit{promote} it and ensure that it is done safely and securely. Access and benefit-sharing for frontier AI could include capacity-building on skills and compute infrastructure and joint priorities for projects aiming to use AI for public goods.

Domestic frontier AI governance may at some point require preapproval for frontier training runs, data centre audits, and independent AI system evaluations throughout training. States may want to harmonise and standardise these regimes with each other and set up systems to share data, information and best practices. This could cover risk thresholds, the details of risk assessments and safety cases, details around post-training evaluations, and information security standards. This development of safeguards can be done by an IAIA. Harmonised standards and rigorous testing can reassure all states that development is secure. 

States would also want assurance that other states are upholding the same level of standards, and not undercutting them. One way to obtain this assurance would be to adopt monitoring and verification mechanisms to ensure that participating countries are not reneging on their agreements to only train AI systems in accordance with the regime’s requirements. The IAIA could audit and monitor compliance with IAIA standards. Participating states could allow the IAIA to monitor these requirements through remote, cryptographic means. In time, they could even allow IAIA auditors to their data centres to run evaluations and tests to see if the facilities and processes comply with the international AI standards. This raises similar questions of information security as the IAEA, and significant further research and pilot projects should be explored in an experimental, iterative way to establish procedures and methods that do not inadvertently reveal or disseminate information. 

An IAIA which serves as a safeguard-developing and monitoring body with a dual mandate of also promoting access and benefit sharing could develop incrementally over time. For example, the IAEA’s first safeguards guidelines in the late 1950s and early 1960s were limited in scope, applying only to research and small power reactors, but they are now extensive and widely respected even with intrusive inspections - 92\% of UN Member States are IAEA Member States (178/193). In the same way, the safeguards and responsibilities of the IAIA could evolve incrementally over time. And again, the IAEA is only one of the most prominent safeguard, monitoring and benefit-sharing organisations - the Organisation for the Prohibition of Chemical Weapons is another. 

\begin{figure}
    \centering
    \includegraphics[width=1\linewidth]{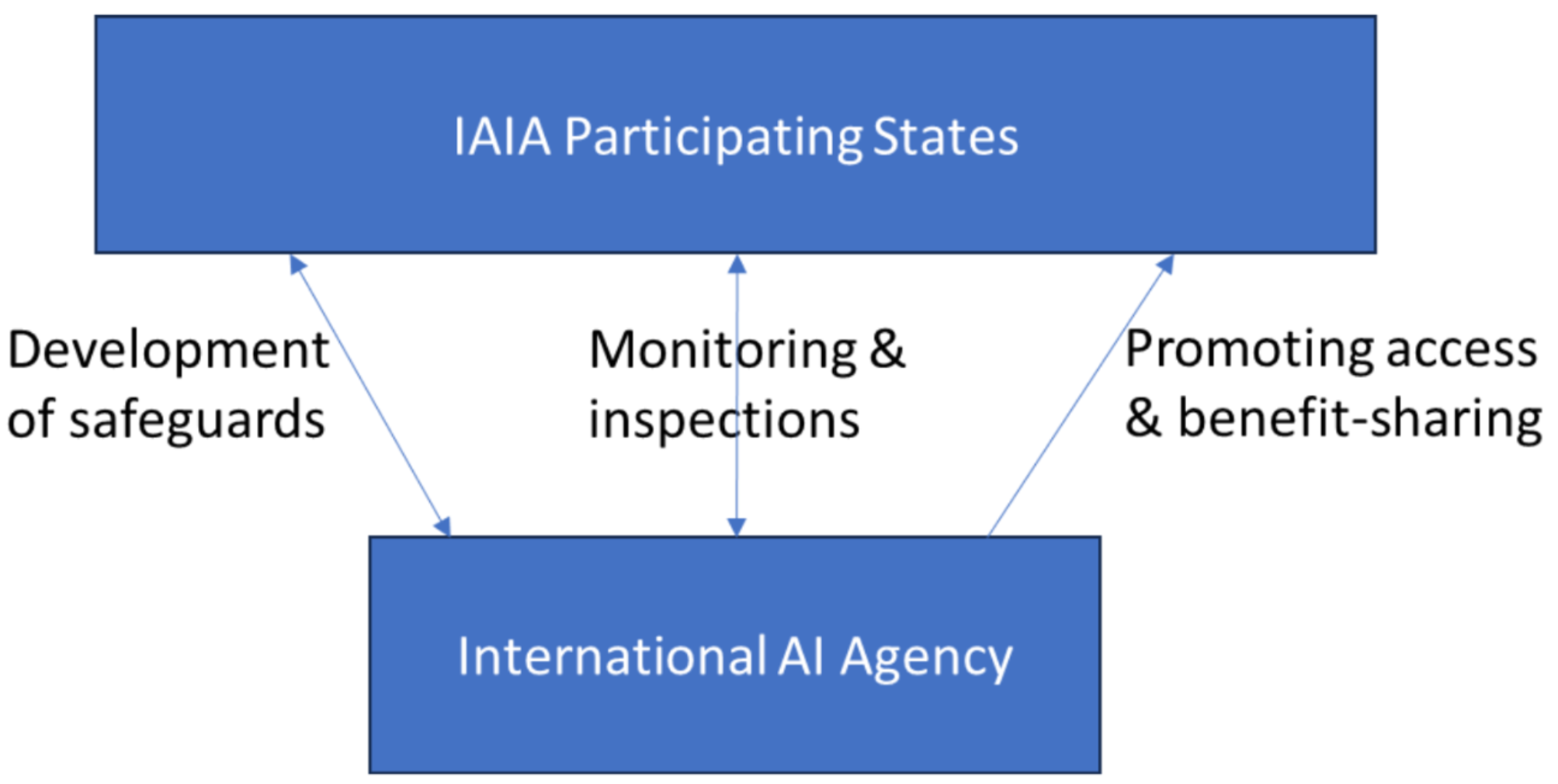}
    \caption{\textbf{The IAIA could carry out three main tasks:} developing safeguard standards, conducting monitoring and inspections, and promoting access and benefit-sharing}
    \label{fig:2}
\end{figure}

\pagebreak
\subsection{Previous discussions of an IAEA for AI}

An IAEA for AI was originally proposed in mid-2023 by several academics, companies, states, and leaders including the Pope and UN Secretary-General (Guterres, 2023; Pullella, 2023; The Elders, 2023). Unfortunately, there was not a substantive positive proposal at the time and the suggestion was criticised (Afina and Lewis, 2023; Mecklin, 2023; Payne, 2023; Roberts et al., 2023; Watson, 2023; Drexel and Depp, 2024). There has been some limited additional discussion since (Stafford and Trager, 2022; Chowdhury, 2023; Hausenloy and Dennis, 2023; Ho et al., 2023; Law and Ho, 2023; Maas and Villalobos, 2023; Markhof, 2023; Baker, 2024; Cha, 2024; O’Keefe, 2024). This section intends to add to that debate and provide a substantive exploration of the case for an IAIA, with a discussion of feasibility and downsides.

There have been three main criticisms of the IAEA analogy: lack of agreement on frontier AI risk, AI being ‘intangible digital software’, and the IAEA itself being a 'failure'. First, the argument is that an IAIA would need significant agreement amongst key states on the scale and nature of the challenge, which existed for nuclear in 1957 but does not for frontier AI today. However, there is a remarkable degree of agreement amongst the key states that frontier AI could pose significant national security challenges. Nevertheless, there is not yet scientific consensus on this open question, and while scientists have a fairly good sense of the risks of this generation of frontier AI, the risks of coming generations are less clear. One key benefit of the evaluations-based approach is that it will produce evidence that answers this question. This will also be contributed to by the AI Summits and Prof Bengio’s State of the Science Reports. That is, rather than speculate on the scale and nature of the challenge, AI evaluations can provide independent, rigorous evidence on the extent to which a given model poses specific risks, for example increasing the biological or cyber capabilities of malicious actors. Second, the argument is that AI is intangible digital software as compared to physical nuclear material. While that may be true of the output, the trained model, it is certainly not true of the process of frontier AI development, which requires hugely expensive equipment and energy. If scaling trends continue, by 2030 frontier training runs will be 10,000 times larger than the GPT-4 training run, using 5GW of energy and tens of millions of H100-equivalents. Analogously to centrifuge enrichment, the training run itself can be monitored and controlled - this is another key benefit of a ex ante regime focussed on the training run. Finally, has the IAEA as a model “already failed” (Watson, 2023)? By contrast, most assessments are that the IAEA has been remarkably successful at developing consensus on safeguards, promoting nuclear energy, and monitoring for non-compliance (Maas \& Villalobos, 2023).

The most similar substantive proposal is for an International AI Organisation (IAIO) (Trager et al., 2023). The relevant analogies are the International Civilian Aviation Organization (ICAO), the International Maritime Organization (IMO), and the Financial Action Task Force (FATF). Broadly an IAIO would monitor and certify \textit{jurisdictions}, whereas an IAIA - like the IAEA - would monitor \textit{projects} (firms and facilities like clusters) and potentially carry out inspections itself. However, these differences are not binary but more of a spectrum. Another key distinction is that the ICAO, IMO and FATF are not backed up by a non-proliferation regime, whereas the IAEA is backed up by the NPT. For more on this see the next section on the Secure Chips Agreement.

\subsection{Next steps to an International AI Agency}

A stylized path to establishing an IAIA could proceed as follows.
\begin{table}
\centering

\begin{tabular}{|>{\raggedright\arraybackslash}p{0.17\linewidth}|>{\raggedright\arraybackslash}p{0.17\linewidth}|>{\raggedright\arraybackslash}p{0.17\linewidth}|>{\raggedright\arraybackslash}p{0.17\linewidth}|>{\raggedright\arraybackslash}p{0.17\linewidth}|}
\hline
\multicolumn{5}{|c|}{International AI Agency} \\
\hline
\textbf{Phase 1} & \textbf{Phase 2} & \textbf{Phase 3} & \textbf{Phase 4} & \textbf{Phase 5} \\
\hline
\textbf{Run first frontier evaluations} & \textbf{Develop the Network of AISIs} & \textbf{Develop safeguards through the Summits} & \textbf{Begin monitoring compliance with safeguards} & \textbf{Trial data centre audits and inspections} \\
\hline
The AISIs will evaluate the GPT-5 generation of frontier models over 2025. & Establish structures for sharing data and information and for collaborative testing and research.& Build common understanding, provide a venue for state announcements and a forum for harmonising domestic regimes.& Compare company publications to regulations and independent findings. Trial cryptographic remote monitoring of compliance.& Audit frontier data centres to ensure that on-chip features (to support monitoring) have not been tampered with. \\
\hline

\end{tabular}
\caption{\textbf{The aims and stages towards an International AI Agency.} Collaboration between the AISIs and harmonisation of domestic regimes through the AI Summits can provide the foundations for an IAIA that sets standards, monitors compliance and promotes access and benefit-sharing.}
\label{Table:1}
\end{table}

\pagebreak
\subsection{Phase 1: Run first frontier evaluations}
Over 2025, the AISIs will run evaluations on the next generation of frontier AI models - likely to include OpenAI’s GPT-5, Google DeepMind’s Gemini 2, Anthropic’s Claude 4, and Meta’s Llama 4. This will provide key evidence as to the scale and nature of security and societal challenges these frontier models pose. One key focus will likely be national security, CBRN and cybersecurity evaluations - to what extent can these models assist malicious users to cause harm? Another will be to track capabilities progress and model characteristics, such as autonomy, longer time-horizon planning, and application to machine learning R\&D itself. Conducting these evaluations will be a key test of the AISIs - to what extent they produce useful, actionable evaluations that are seen as a real value-add by governments and companies.

The results of these evaluations and testing will likely be shared at the very least confidentially with partners and likely publicly - though perhaps with some confidential details redacted or anonymised. Two key venues for sharing these results will be the meetings of the Network of AISIs and the AI Summits. 

\subsection{Phase 2: Develop the International Network of AISIs}

The practical experience and details of running these evaluations will likely highlight areas where harmonisation and standardisation would be useful. One avenue forward for this will be collaboration between the AISIs, such as the first meeting of the International Network of AISIs in November 2024. Collaboration could build on existing Memoranda of Understanding. For example, the UK and US AISIs agreed to information-sharing, close cooperation, expert personnel exchanges and to:
\begin{quote}
“build a common approach to AI safety testing and to share their capabilities [...] They intend to perform at least one joint testing exercise on a publicly accessible model. They also intend to tap into a collective pool of expertise by exploring personnel exchanges between the Institutes.” (DSIT, 2024d)
\end{quote}
AISIs are likely to want to be able to share information (such as “Company X told us that they intend to begin training on the 1st September”) and data such as datasets of potentially dangerous responses to prompts, or interpretability data. These AISIs will need to set up systems to share this information and data in secure ways. This kind of AISI collaboration will be in AI companies’ interests too. They will not want to provide access in lots of different, tailored ways; provide the same information repeatedly; or wait to go through many rounds of evaluations that provide the same evidence. They would prefer some degree of harmonisation between the AISIs.

Once the evaluations are carried out on the next generation of frontier models, the AISIs are likely to branch out and conduct other kinds of AI safety research, as well as evaluating other AI models such as ‘biological design tools’ (BDT) (Sandbrink, 2023). The AISIs will also need to assess, iterate and improve their evaluation processes for the following generation of frontier models. This could also involve some degree of specialisation between the AISIs - for example one could press forward with fundamental research and concrete evaluations around mechanistic interpretability, while another focussed on biological upskilling. This could be more efficient than each investing a small amount into the entire range of concerns.

This collaboration can lay the foundations for shared IAIA safeguards and monitoring. If the AISIs agree ‘the best practice for evaluations on topic X is to take steps A-Z’ and then implement them in their domestic regulatory regimes, that would be a standard. This can be seen, for example, as highly similar to IAEA safeguards for assessing nuclear safety. 

The meetings, joint work and information sharing of the AISIs are likely to be in-depth and technical. More general policy matters are likely to be agreed at the AI Summits.

\subsection{Phase 3: Develop safeguards through the AI Summits}

The AI Summits serve three purposes: sharing evidence and creating common understanding across different international divides; providing a venue and ‘hook’ for state announcements; and providing a forum for harmonising domestic regimes and making shared agreements between companies and governments.

First, one purpose is sharing evidence and establishing common understanding of AI capabilities and risks. At Bletchley, the UK Government released a report on ‘Capabilities and risks from frontier AI’ (DSIT, 2023). At Seoul, the interim ‘State of the Science’ report was published, led by Prof. Yoshua Bengio, one of the world’s most respected AI scientists (Bengio et al., 2024), with the full report being published at the France AI Summit. Some of the responsibility for writing an IPCC-style report is likely to shift to the United Nations, following agreement of the Global Digital Compact at the September 2024 UN Summit of the Future (Pouget et al., 2024). Nevertheless, the AI Summits will remain a key venue for the regular production of reports and sharing of information. This is a key part of the IAEA’s work too. As a forum, this AI Summit series has had high-level and productive participation from China. The Summit series have had twin tracks - one larger including many countries and China, and one smaller and focussed on US allies. This allows for sharing information and generating common understandings, while also advancing some more detailed agreements.  

Second, states will also use the occasion of the regular AI Summits to highlight their particular priorities. Hosts can shape the program and attendees can use the Summits as a ‘hook’ to announce their own initiatives. For example, the US launched its Executive Order just before the Bletchley Summit. The France AI Summit's five workstreams cover AI public goods; the labour market; culture; trust, safety and security; and global governance. Those states that have particular expertise and interest in AI security can progress this forward under the ‘AI of trust’ workstream. Summit head Anne Bouverot called for three initiatives to be launched at the French AI Summit (in the recommendations of the French National AI Commission): a €500 million International Fund for Public Interest AI; a "1\% AI" solidarity mechanism for funding computing infrastructure in developing countries; and an Organisation mondiale de l’IA (World AI Organization) to define binding standards for AI systems and their auditing. So it is clear that France is emphasising access and benefit-sharing at their ‘AI Action Summit’. This is an important focus and a key part of any IAIA’s ‘dual mandate’. It builds on efforts to include geographic diversity in the attending states of previous Summits. 

Third, as agreed in Seoul, one key next step for the AI Summit Series is for states to define risk thresholds and agree on appropriate responses to different thresholds: 
\begin{quote}
“We plan to collaborate with the private sector, civil society and academia, to identify thresholds at which the level of risk posed by the design, development, deployment and use of frontier AI models or systems would be severe absent appropriate mitigations, and to define frontier AI model or system capabilities that could pose severe risks, with the ambition of developing proposals for consideration in advance of the AI Action Summit in France.” (DSIT, 2024c)    
\end{quote}
Sixteen AI companies also committed to publish their policies on ‘responsible capability scaling’ ahead of the France AI Summit, detailing their views of capability risk thresholds and what steps they commit to take at those different levels. Those that have been published discuss particular levels of capabilities and risks, and detail requirements at those different ‘AI Safety Levels’ (ASL), by analogy to biosafety levels (BSL).

States agreed to work towards a shared understanding and a level of agreement on thresholds and best practice when those thresholds are hit. At future Summits these agreements could deepen and become more explicit, and even be harmonised as standards (analogous to IAEA safeguards). For example, most companies and states agree that information security is important to prevent model theft and misuse by criminals. There is also emerging agreement that this information security should increase as capabilities - and therefore the risk of theft and misuse - grows. This has been formalised as RAND information security levels SL1 to SL5 that may be appropriate for different ASLs (Nevo et al., 2024). This can be harmonised and agreed upon at the Summits. Harmonising the requirements of domestic regulation establishes the foundations for safeguards and monitoring.

This set of agreements can be viewed as analogous to the initial safeguards set by the fledgling IAEA. The IAEA analogy would be that in 1956/57 the first full scale nuclear power plants were being built. There were domestic nuclear safety and security regimes, but details were still being worked out. Harmonisation and standardisation on safeguards amongst domestic regimes helped regulators adopt best practice, helped gain public support for nuclear energy, and helped provide stability and confidence to the industry. Analogously, we are still at an early stage with frontier models. There are emerging domestic frontier regimes but details are still being worked out. Harmonisation on frontier AI safeguards could help domestic regulators, contribute to public trust, and provide stability and confidence to the industry. The Summit series is a natural venue to work towards those agreements.

It could also be useful to have a permanent secretariat for this Summit series - at the moment the secretariat is provided by whichever government is hosting each Summit. A secretariat would still work closely and be led by the organising government, but also provide continuity and expertise. A secretariat could, for example, work with participating states between the Summits to take forward discussions on work agreed at previous Summits and it could follow up with the companies to ensure they are keeping to their frontier AI safety commitments. An independent secretariat could provide assurance to participating states of continuity and progress.

\subsection{Phase 4: Begin monitoring compliance with safeguards}

In order to ensure that states and companies are abiding by the commitments they have made, it could be helpful to have monitoring and verification mechanisms. This would reassure all states that they are not being exploited by other states cheating, and that their national security is not being threatened by insecure development in other states. Similarly, the IAEA monitors the safety and security of facilities like nuclear power plants to ensure that firms and states are complying with agreed-upon safeguards.

This monitoring can be done in unintrusive ways. The easiest is checking that companies have indeed published documents and information they have promised to release, such as responsible capability scaling policies. These responses can be compared to each other and to best practice (LCFI, 2023). They can also be compared to other company policies, domestic regulation and findings from independent auditors like the AISIs. For example, if the companies responsible capability scaling policy mandates a particular cybersecurity level at ASL-3 or ASL-4, one can check whether the information security was indeed at that level, as it was agreed it would be. If a frontier AI system is released, was it notified in the relevant jurisdiction and did the company share risk assessments and evaluation access to third-party assessors? The discontinuous nature of frontier AI development helps this - rather than monitoring lots of little releases, most of the action is in big jumps to a new frontier generation every one to two years. Like domestic regulatory regimes, this means a low regulatory burden. 

Over time, other forms of monitoring and verification could provide more reassurance. For example, my co-authors and I have proposed monitoring solutions such as 'proof-of-work' (Sastry et al., 2024). This kind of remote monitoring leveraging cryptographic techniques that do not reveal the content of the computation can be fairly light-touch while also providing reassurance.

\subsection{Phase 5: Trial data centre audits and inspections}

In fulfilling its monitoring and verification obligations, the IAEA now complements information-sharing from countries with on-site inspections (e.g. of nuclear power plants) and interviews with scientists and employees. This provides additional reassurance that IAEA standards are being complied with. In arms control, this level of monitoring and verification is seen as the gold standard. That could be an ultimate aspiration for the IAIA. However, this is clearly a long-term goal - the IAEA’s ‘INFCIRC/153’ model agreement on inspections was only agreed in the early 1970s, thirteen years after its founding. 

On-site inspections can be politically contentious. Soviet objections to on-site inspections proved a stumbling block to arms control agreements during the Cold War. When agreements were eventually made in the early 1970s, monitoring and verification was conducted by “national technical means”: satellites, overflight photography, seismographs and intelligence-gathering. This was unilateral and did not require cooperation. In the 1990s, the US pharmaceutical industry lobbied against a Verification Protocol to the Biological Weapons Convention over concerns about losing trade secrets. Any inspections of data centres should be focussed only on the largest clusters and should be carefully designed so they do not undermine trade secrets. For example, inspections could be limited to inspecting randomly chosen chips/racks to see if the remote monitoring capabilities had been tampered with. This would be greatly assisted by the development of unique chip IDs and on-chip hardware features (on which see next section).

Should an IAIA include China? It is important that democratic allies are not held back by complying with safeguards and monitoring that others ignore. If all the main states where frontier AI is being developed are all included in the same international forum, it could reduce the risk of undercutting on security and safety. However, three major concerns could scupper this: general antagonism and the ‘new Cold War’ between the USA and China, specific Chinese objections to the US-led chip export controls, and concerns over Chinese espionage from the USA and its allies. More likely then is an IAIA composed of US allies, without China. One advantage of this approach is that participation in the IAIA could be encouraged by another multilateral regime composed of US allies which excludes China: the chip export controls. I discuss that in the next section on the Secure Chips Agreement.

One eventual shared goal of the twin tracks of international collaboration between the Network of AISIs and the AI Summits could be an International AI Agency that develops safeguard standards, promotes access and benefit-sharing, and conducts monitoring and inspections.

\pagebreak

\section{Secure Chips Agreement - a Non-Proliferation Treaty for state-of-the-art AI chips}

I now explore the next steps for the AI export controls regime. How can a non-proliferation regime for state-of-the-art chips help ensure that frontier AI systems are not trained in states where they may be more likely to be unsafe and insecure? 

The beginnings of a de facto non-proliferation regime for state-of-the-art AI chips came with the export controls on advanced chips in October 2022. Current export controls affect the entire global market. They are intended to curb proliferation of certain chips and therefore curb misuse of AI - especially by non-allied states. The main aim is to limit adversaries’, in particular China’s, ability to buy or build high-end, data-centre AI chips - which I will refer to as state-of-the-art chips. It has been mostly successful in that aim (Fist and Grunewald, 2023). The regime began as largely unilateral and has subsequently been joined by other US allies. These export controls rely on the cooperation of key compute producer states like Japan, the Netherlands, Taiwan, and South Korea. Those states currently see participation as in their interest due to factors such as shared concern at Chinese capabilities and the alliance politics of partnering with the USA. The U.S. and allies are coordinating on strong export controls on both state-of-the-art AI chips and advanced AI chip inputs, such as operational fabs, and the few key chokepoints in the semiconductor supply chain, such as wafers, EDA software, EUV lithography machines, and advanced optics. They are also monitoring the regime to assess the extent of diversion/smuggling and planning more enforcement at a domestic (United States Department of Justice Office of Public Affairs, 2023) and international level (through, e.g., end-use checks).

The export controls regime will likely continue to be incrementally strengthened over the coming years to ensure the continued nonproliferation of high-end AI chip inputs and AI chips. Options include assigning unique IDs to state-of-the-art chips and tracking them in a registry; implementing chain-of-custody and audits; and on-chip features that could enable location verification, secure enclaves, networking limits, multiparty control and remote enforcement. 

However, the export control regime also provides the foundation for achieving other goals. It could be extended to further incentivize the safe and secure development of AI. This section outlines an international governance scheme called a “Secure Chips Agreement” - an ‘NPT for AI’. A non-proliferation agreement can reinforce, support and provide incentives for participation in an international institution - and vice versa. A decade after the IAEA was established in 1957, it was reinforced by 1967’s Treaty on the Non-Proliferation of Nuclear Weapons (more commonly known as the Non-Proliferation Treaty or NPT), a regime for the international control of nuclear material. Article III of the NPT restricts parties from exporting nuclear material or processing equipment to states which have not been determined to be compliant with IAEA safeguards. The IAEA shapes those safeguards and serves as a third-party auditor, collecting information and carrying out inspections to monitor compliance. Analogously to the NPT, the Secure Chips Agreement would be an international agreement by participating countries to restrict parties from exporting (large quantities of) state-of-the-art AI chips to states which have not been determined to be compliant with IAIA safeguards. The IAIA could shape those safeguards and help monitor compliance. The assurances provided by the IAIA could help enable the continued secure transfer of advanced AI chips and therefore enable safe and secure frontier training runs, promoting the dual mandate of the IAIA.\footnote{Other analogous multilateral commitments include multilateral export control regimes: the Wassenaar Arrangement, Nuclear Suppliers Group, Australia Group and the Missile Technology Control Regime; the informal Zangger Committee; and the defunct Coordinating Committee for Multilateral Export Controls. Other regimes that can provide inspiration and analogies include the monitoring and verification provisions of the bilateral New START Treaty or the multilateral Organisation for the Prohibition of Chemical Weapons (OPCW).} 
\begin{table}
\centering

\begin{tabular}{| l  |>{\raggedright\arraybackslash}p{0.45\linewidth}|l |}
\hline
 & \textbf{Safeguards and Monitoring body} & \textbf{Non-Proliferation agreement} \\
\hline
\textbf{Nuclear} & International Atomic Energy Agency & Non-Proliferation Treaty \\
\hline
\textbf{Frontier AI} & International Artificial Intelligence Agency & Secure Chips Agreement \\
\hline

\end{tabular}
\caption{\textbf{For both nuclear and frontier AI, an agency that can develop safeguards and provide monitoring and an agreement on non-proliferation
can reinforce one another.}}
\label{table:2}
\end{table}

The Agreement would operate by restricting non-participating states’ access to the most advanced state-of-the-art AI chips. Participating states would be required to apply domestic policies like aligning with IAIA safeguards and tracking their state-of-the-art AI chips. This builds on the previous two sections: this is an internationalization of the domestic regime and a reinforcement of international oversight - it combines domestic frontier regulation, the IAIA and the export control regime. Countries that sign a verifiable commitment to certain safe and secure chips practices and are determined to be compliant are then permitted to accumulate large amounts of compute via purchases of state-of-the-art AI chips. Those that are not determined to be compliant are not permitted. This does not have to be permanent - if the country comes into compliance, the participating states can reassess.

Crucially, state-of-the-art chips are hard to stockpile. Technical improvements happen so quickly that previous generations are quickly out of date in a matter of years or even months. Companies need to continue purchasing new generations of chips to continue being competitive - withholding that ongoing access makes them unable to compete at the frontier. 

As discussed above, key compute producer states currently participate in the current export controls due to shared concern at Chinese capabilities and US alliance politics. States may wish to participate in the Secure Compute Agreement for similar reasons, bolstered in coming years by other concerns. Most importantly, shared concern about frontier risk is likely to prompt them to establish frontier AI regulatory regimes, as they do not want frontier AI systems developed in a potentially insecure manner without their oversight. They would not want frontier training runs to occur in other states without these safeguards - as the potentially unsafe, insecure frontier AI systems could damage them. More narrowly, if these states have put in place frontier AI regulatory regimes, they will not want those regimes and their economic competitiveness undermined by freeriding. Those states will not tolerate some other state which plans to set itself up as an insecure ‘compute haven’ (analogous to a tax haven). Why would they export state-of-the-art AI chips to such a compute haven which would undercut their security, regulation and competitiveness? Finally, whether they initially agreed with it or not, once a non-proliferation regime is in place, states would have an incentive to participate due to their interest in retaining access to advanced AI chips.

\begin{figure}
    \centering
    \includegraphics[width=1\linewidth]{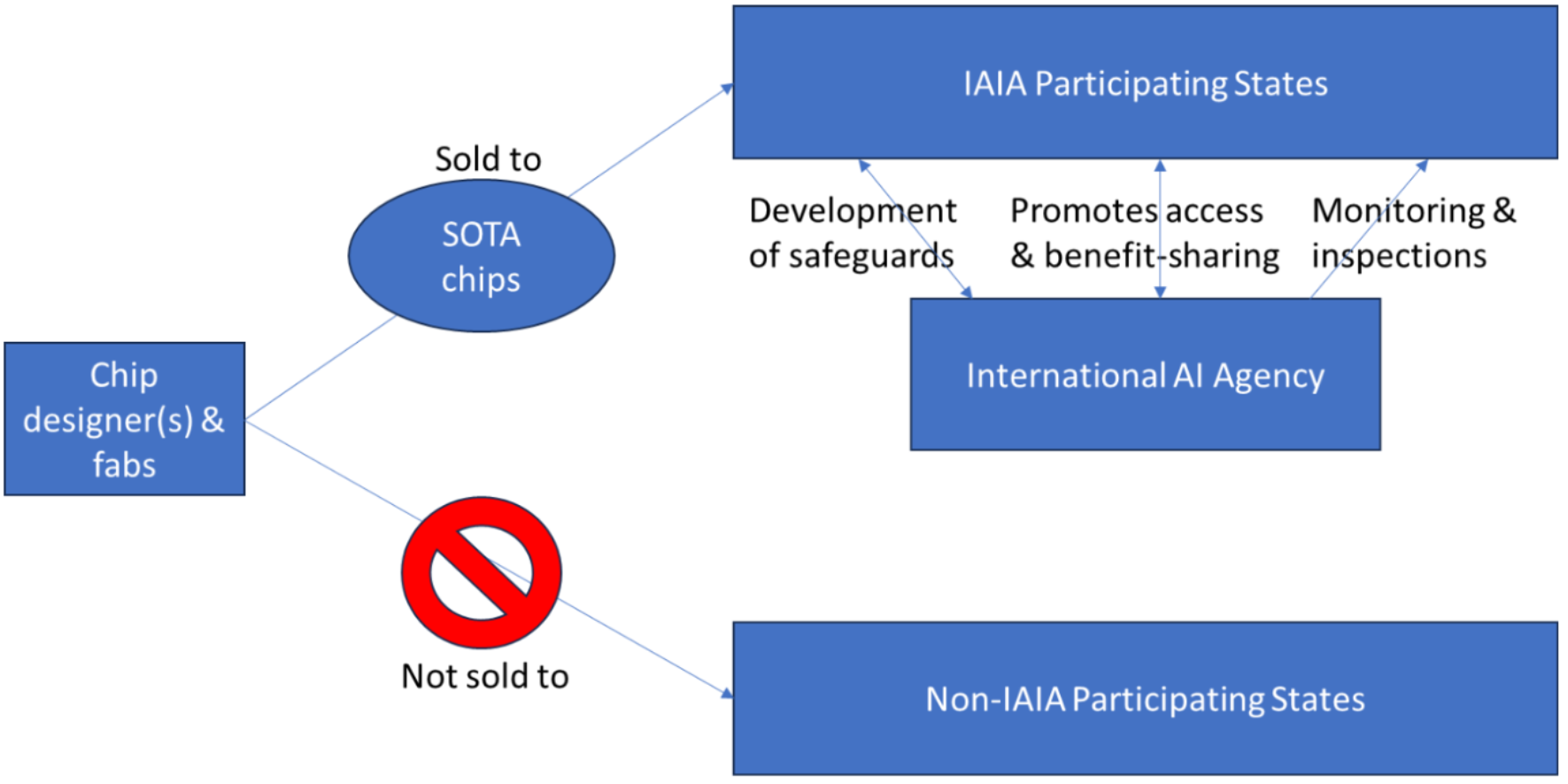}
    \caption{\textbf{The carrot-and-stick of access to state-of-the-art chips.} States that participate in the IAIA and are determined to be compliant are able to import state-of-the-art chips, those that are not determined to be compliant are not allowed to do so. This is an extension of the existing state-of-the-art chips non-proliferation regime.}
    \label{fig:4}
\end{figure} 

It is important to note that this should only apply to state-of-the-art AI chips and the very advanced equipment needed to produce them. As I have repeatedly emphasised, small-scale compute and non-AI compute should mostly be excluded from on compute governance. With my co-authors, we estimated that state-of-the-art chips represent less than 1\% of all high-end (less than 7nm) chips, and less than 1 in 400,000 (0.00026\%) of all chips (Sastry et al., 2024, p. 67). To adapt US National Security Advisor Jake Sullivan’s term, this is a high fence but a very small yard.

The NPT is a key inspiration for the Secure Chips Agreement. The analogy between AI training and uranium enrichment is highly suggestive. Advanced data-centre quality AI chips are rare and costly. Indeed, arguably the supply chain for state-of-the-art chips is more concentrated and therefore easier to monitor and secure than that for nuclear/fissile material: for each of the dual-use goods under the Nuclear Suppliers Group’s purview there are 6-59 manufacturers (Doyle 2019), while many steps in the compute supply chain have only a single company (Sastry et al., 2024). Moreover, those companies are concentrated in a handful of countries, all of them allies: the USA, Taiwan, the Netherlands, Japan and South Korea. So the Secure Chips Agreement only needs the active participation of a small group. This supply chain concentration seems unlikely to change for the foreseeable future - it was built up over decades at the cost of trillions of dollars. That there are several near-monopolies with high margins despite attempts by other companies to compete shows how high the barriers to entry are.

\begin{figure}
    \centering
    \includegraphics[width=1\linewidth]{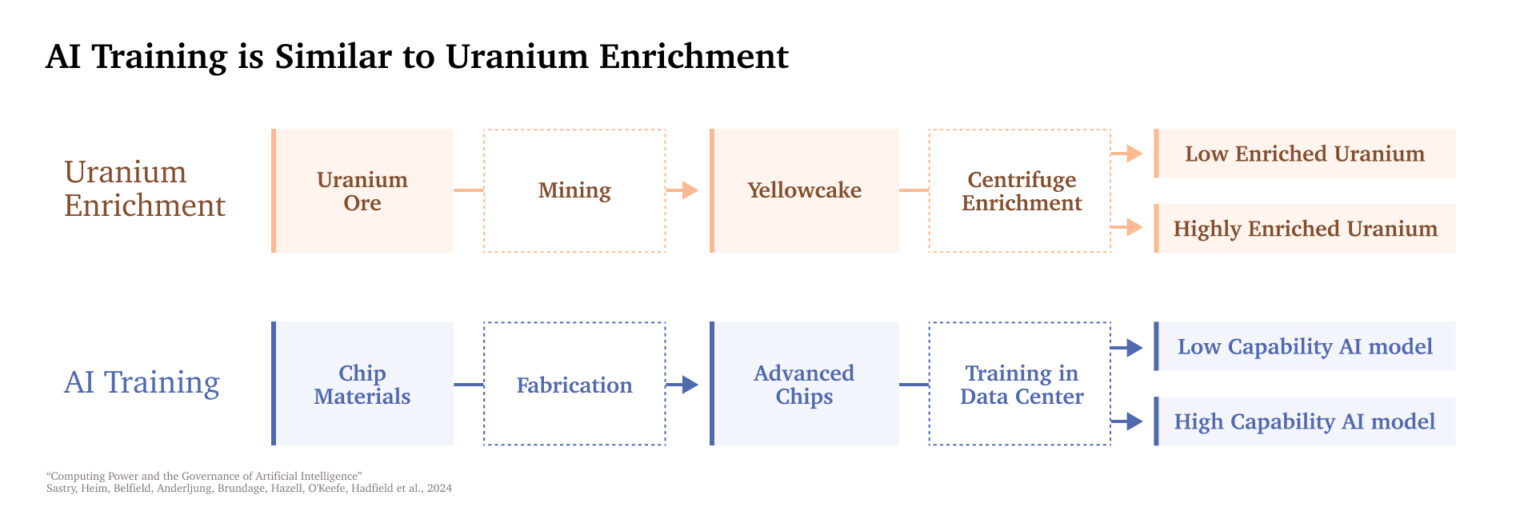}
    \caption{\textbf{The analogy between uranium enrichment and AI training.} For both AI (chips) and nuclear energy (uranium), there is a key input that is difficult to produce and potentially regulable.}
    \label{fig:5}
\end{figure}

\begin{figure}
    \centering
    \includegraphics[width=1\linewidth]{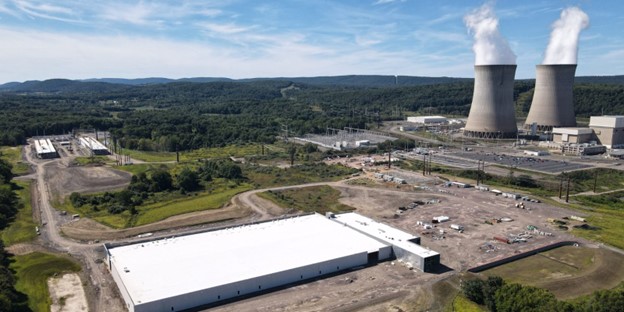}
    \caption{\textbf{AWS’ nuclear-powered 960-MW data centre campus.} Both sides of this picture - AI training data centre and nuclear power plant - are detectable and monitorable. Photo: Talen Energy.}
    \label{fig:6}
\end{figure}

There are clear downside risks to such a proposal, such as diversion incentives, hierarchy, and tension with China. First, a non-proliferation regime incentivises attempts to circumvent these controls, such as divergence/smuggling, low-compute algorithmic progress and indigenous compute supply chains. However, these incentives exist with the current export control regimes, and most assessments are that while there is some diversion, it is relatively small and insignificant and despite substantial investment indigeneity is years if not decades away. Second, a Secure Chips Agreement may be criticized for similar reasons to the NPT: as setting up an unfair divide between those states that may and those that may not possess a dangerous capability, and as being arrogant overreaching (Egeland, 2017). Large importers of advanced AI chips who nevertheless do not yet currently have an effective frontier AI regulatory regime such as the United Arab Emirates or Saudi Arabia could chafe if their access to advanced AI chips is limited. Third, it could be viewed as additionally antagonistic with China, though this would not be a completely fair assessment: the current export control regime cuts off all access to advanced chips, while the Secure Chips Agreement at least provides a path for loosening this restriction if assurances can be verified. Any critique of the Secure Chips Agreement proposal must reckon with the reality that the chip export controls now affect the worldwide market, and have such bipartisan US support that they will continue for the foreseeable future.

Policy-makers do not have to decide to commit to the IAIA and Secure Chips Agreement any time soon. Progress on domestic frontier regulation, harmonisation and standard-setting, and monitoring will take time. No matter what, the US and allies are highly likely to take steps over the coming years to strengthen and make more sophisticated the current export controls regime. Doing so preserves the option of using this chokehold on compute to reinforce the IAIA in coming years.

Obviously there is much more to be said for and against this proposal, and significantly more research and discussion should be conducted before decisions are made. This is a major debate, and one I seek to contribute to with this paper, rather than attempt to resolve. 

\subsection{Next steps to a Secure Chips Agreement}

The current export control regime could be deepened and strengthened to ensure continued non-proliferation. The initial goal would be to maintain the current allied states’ leverage over the most advanced computing hardware relevant for AI. Eventually, this could be turned into a Secure Chips Agreement to reinforce the harmonised standards around domestic frontier AI regulation. A stylized path to a Secure Chips Agreement could proceed as follows.

\subsection{Phase 1: Unique IDs \& chip registry for state-of-the-art chips}

For the U.S. and its allies to ensure continued nonproliferation of chips and to prevent diversion to countries and customers who should not have access, these countries could establish unique IDs and a registry for tracking the transfers and ownership of state-of-the-art chips along their lifecycle - the flow and stock of these chips from fabrication to data centre to destruction (Fist and Grunewald, 2023; Grunewald and Aird, 2023; Shavit, 2023; Cheng, 2024; Grunewald and Fist, 2024; Sastry et al., 2024). Chip producers, sellers and resellers would report transfers which would be logged in a registry. At a minimum, a registry could be established with reports from only six companies: NVIDIA, TSMC, Alphabet, Amazon, Microsoft and Meta. These companies are sophisticated market players fully capable of tracking and reporting these transfers. This chip tracking could be extended to the chip level: each state-of-the-art chip could have a unique digital ID, or this could be physically added to the state-of-the-art chip during production.

In parallel with establishing global AI chip tracking, these states could also collect and share some limited, aggregated information about data centres and cloud computing jobs in their jurisdiction (Sastry et al., 2024). Large compute providers are required to notify the US government both if they own large data centres with state-of-the-art chips, and also to notify about large cloud computing jobs. In both cases, the US Government wants to know who can use large concentrations of state-of-the-art chips. If frontier training runs could pose significant national security risks, then it makes sense to enlist those hosting such computations - cloud compute providers - to manage and mitigate such risks (Heim et al., 2024). A first step could be to convene substantive technical working groups drawing on corporate, government and other expertise to begin discussing data centre reporting requirements and how to share such reporting with each other. 

\pagebreak

\begin{table}
\centering

\begin{tabular}{|>{\raggedright\arraybackslash}p{0.22\linewidth}|>{\raggedright\arraybackslash}p{0.22\linewidth}|>{\raggedright\arraybackslash}p{0.22\linewidth}|>{\raggedright\arraybackslash}p{0.22\linewidth}|}
\hline
\multicolumn{4}{|c|}{Secure Chips Agreement} \\
\hline
\textbf{Phase 1} & \textbf{Phase 2} & \textbf{Phase 3} & \textbf{Phase 4} \\
\hline
\textbf{Unique IDs \& chip registry for tracking state-of-the-art chips} & \textbf{Chain of custody and random audits} & \textbf{On-chip hardware features} & \textbf{Extend to reinforce IAIA safeguards and monitoring} \\
\hline
This includes setting up unique IDs and a registry for tracking state-of-the-art AI chips throughout their lifecycle. & Compute suppliers implement ‘chain of custody’ for state-of-the-art AI chips. This phase may involve physical audits of suppliers to ensure chips are not being diverted. & State-of-the-art AI chips are required to have hardware security features, which could include location verification, secure enclaves, networking limits, multiparty control and remote enforcement. & State-of-the-art AI chips are withheld from non-participating states. If a state is not participating in the IAIA, then it would not be allowed to accumulate a large quantity of state-of-the-art AI chips.\\
\hline

\end{tabular}
\caption{\textbf{The aims and stages of achieving a Secure Chips Agreement. }This is an international agreement that prohibits members using their state-of-the-art AI chips on excessively risky AI training runs. It aims to achieve nonproliferation of risky AI models while incentivizing secure AI development. It operates by restricting non-participating countries’ access to the most advanced AI chips.}
\label{table:4}
\end{table}

\subsection{Phase 2: Chain of custody and random audits}

In this phase, chip tracking from the previous steps could be augmented to implement chain of custody. Chain of custody is a paper trail of who has responsibility for a state-of-the-art chip. If a state-of-the-art chip is diverted, responsibility is assigned to the latest holder in the chain of custody (Cheng, 2024). This helps assign responsibility for diversion, providing strong incentives for each company in the chain to detect and discourage diversion. 

The register could be randomly audited to detect and assign responsibility for diversion. In-person inspections and physical audits of different suppliers in the value chain - from fabs to data centres - could improve assurance that the registry is working and state-of-the-art chips are not being diverted.  It would not be necessary to audit every single chip or every single seller: a data centre capable of conducting a frontier training run needs a lot of state-of-the-art chips. Random auditing would probably be enough to catch evidence of diversion.

\subsection{Phase 3: On-chip hardware features}

Because the production of the latest AI chips takes place entirely within the Secure Chips Agreement states, they could also require that these chips be produced with hardware security features that make it difficult for AI models to be stolen.  Hardware security features such as secure enclaves (also known as Trusted Execution Environments) are common on commodity chips, but are less common on state-of-the-art chips (Brundage et al., 2018, 2020; Avin et al., 2021). For example, the latest chips might have hardened dedicated security modules, which help avoid attacks like tampering, spoofing, and theft - or to support other on-chip features, such as remote attestation. State-of-the-art AI chips could be required to have other hardware features that assist non-proliferation, which could include location verification, networking limits, multiparty control and remote enforcement (Aarne, Fist and Withers, 2024; Sastry et al., 2024).

Remote location verification can assist an audited registry (Shavit, 2023; Brass, 2024). The best developed concept is for delay-based geolocation: a cryptographic challenge is sent to the chip and the time the chip takes to respond is measured. If the chip is physically located where it is supposed to be, it will take the expected time; if it is not then it will take longer. For example, if a chip had been diverted to China, it would take a few microseconds longer to respond than if it were in Europe where it was supposed to be. This is harder to spoof than asset-reported GPS or RFID tracking or topology-based geolocation using public internet infrastructures (IP addresses etc).

Networking limits involves limiting the bandwidth of chip-to-chip networking - this prevents a user connecting them up into a large cluster needed for frontier training. It is a usage restriction rather than a way to better track chips. It would not directly detect diversion, but it would discourage it and prevent diverted state-of-the-art chips being used in a large cluster. It could also allow for more fine-grained export controls: network-limited chips could be exported to states to which participating states would not export normal chips.

Remote enforcement involves a specialised co-processor on the chip requiring an ongoing cryptographically signed certificate from a counterparty. Without this ongoing approval, the chip’s performance would degrade or it would cease working. If a batch of chips were diverted the signal could and would be switched off, discouraging diversion.

Multiparty control involves distributing the ability to begin a frontier training run through multisignature cryptographic protocols. The chips for the training run would only begin working once the workload had been cryptographically approved by all parties. This is akin to distributing a veto. Again, this discourages diversion - without that approval the diverted chip is not useful.

On-chip hardware features take time to research, develop and incorporate into state-of-the-art chips in ways that do not cost too much or harm the efficiency or security of the chip. This is why it is the third phase - the technical R\&D process to develop these features has begun already, and the results can be incorporated in coming generations of state-of-the-art chips. Participating states can convene technical experts to revise and extend these safety and security features, while continuing to restrict access to state-of-the-art chips.

\subsection{Phase 4: Extend Secure Chips Agreement to reinforce IAIA safeguards and monitoring}

Maintaining and strengthening the export controls preserves the option for participating states to use their control of the chip supply chain to achieve other objectives. A third party country setting itself up as an unregulated ‘compute haven’ would pose serious challenges to Secure Chips Agreement states’ security and economic competitiveness. These states could withhold access to state-of-the-art chips - and thus the most powerful AI systems - from such compute havens.

Over the past few years the export control states have woken up to the fact that they control the supply of one of the most important products in the world. If they so choose, they can use their control of this bottleneck to further discourage unsafe and insecure AI development.

\pagebreak

\section{A US-led Allied Public-Private Partnership for Frontier AI}

How can governments be assured that frontier AI is being developed by a project that is legitimate, safe, secure, non-adversarial, peaceful, not prone to misuse and feasible?

In this section I explore a “US-led Allied Public-Private Partnership for Frontier AI”: an international megaproject between the US Government, its allies and leading Western AI companies. By US allies I mean liberal democracies such as the UK and EU (NATO allies) in Europe; Japan, South Korea and Taiwan in East Asia; and Canada and Australia (Five Eyes allies). Key companies could include NVIDIA, ASML, TSMC, Alphabet, Amazon, Microsoft, Meta, OpenAI and Anthropic. 

This idea has often been described as ‘CERN for AI’, drawing inspiration from previous international scientific megaprojects: the European Organization for Nuclear Research (CERN), the International Space Station (ISS), and the International Thermonuclear Experimental Reactor (ITER). All three are notable as collaborations between several states - including geopolitical adversaries. An AI megaproject could have different possible objectives. Alternatives include a ‘CERN for AI’ in general, a ‘CERN for AI for Good’, or a ‘CERN for AI Safety’. In this section I discuss a project “for Frontier AI”: one that builds large-scale compute clusters and carries out frontier training runs on the path to AGI. To date there has been no systematic evaluation of - or significant interstate dialogue on - the details of such an international AI megaproject, though see (Fischer and Wenger, 2019; Kemp et al., 2019; Kerry, Meltzer and Renda, 2022; Stix, 2022; Zhang et al., 2022; Hausenloy, Miotti and Dennis, 2023). This section builds especially on my and my co-authors’ analysis in ‘Collaborating on a joint AI megaproject’ (Sastry et al., 2024, p. 51). 

The CERN analogy suggests the potential value of joint investment in capital-intensive infrastructure. The development of CERN was driven by the increasing - and daunting - high fixed capital costs of particle accelerators. Beyond a certain point, the size, and thus the energy, of the accelerator was the key to making progress. The costs quickly stretched beyond the budgets of individual scientists or groups, then beyond the budgets of individual universities, then beyond the budgets of small European countries. The alternative was to pool funding to build the capital-intensive, expensive, specialized, joint infrastructure for scientific experiments. Frontier AI also faces increasing, daunting high fixed capital costs of building AI infrastructure. Beyond a certain point, the size of the training run - and the compute cluster to run it on - has become a key component of making progress at the frontier. These costs are beyond the budgets of academia. In the case of AI, the capital-intensive infrastructure that states would jointly invest in is likely to be extremely large compute clusters.

However, there is a serious disanalogy with CERN, ITER or ISS. These three are primarily scientific megaprojects without significant dual-use implications, which include adversaries and for which scientific considerations are intended to be determinative (Barish, 2013; Gagnon, 2016; Robinson, 2019, 2021). Frontier AI is significantly more dual-use, and has much greater national and international security implications - indeed it is arguably a vital strategic asset (Gaithersburg Md Nist, 2024; Hickey, 2024). This would make cooperation on such infrastructure between adversaries significantly more fraught than it is for CERN. Moreover, the risks of espionage and model theft are very high. A joint project - a collaboration between the USA and China -  is neither feasible nor desirable.

For this reason, I will refer to a US-led Allied Public-Private Partnership for Frontier AI. A more relevant analogy is the Joint Strike Fighter Program to develop the F-35 fighter. This is structured as a public-private partnership with Lockheed Martin and sub-contractors. This is a US-led project, but with the participation of US Allies (UK, Canada, Australia, Netherlands, Denmark, Norway), sharing costs and providing resources. There are three ‘Tiers’ which reflect financial contributions, extent of technology transfer and ability to bid on subcontracts, and order of access. The initial R\&D cost was \$400bn (with lifetime costs in the trillions) - which is comparable to the cost of building and operating a data centre cluster capable of a \(10^{29}\) training run. It is also worth bearing in mind that the JSF program has become notorious for cost and timeline overruns.

There are other defence analogies one could draw on, for example the French Charles de Gaulle air-craft carrier uses a US-designed and built steam catapult; and British nuclear submarines used Polaris and now use Trident missiles, both US systems. Other more civilian yet still dual-use analogies could include the CFM International CFM56 engines (built by France’s Safran and USA’s GE Aerospace, used by Airbus and Boeing) and Intelsat (a European and US collaboration). There are also often collaborations and joint ventures in pharmaceuticals or oil and gas. Possible analogies of projects that began in US defence and then became civilian and broadly used are the internet and GPS.\footnote{European analogies (not involving the USA) could include Airbus (a France-Germany-Spain collaboration), Galileo (the EU equivalent of GPS), and the Ariane rockets built for the European Space Agency.} 

There are several national supercomputing projects amongst US allies. The U.S has long standing supercomputing programs, such as the Department of Energy’s supercomputing office. These have often been used for nuclear research. More clearly civilian projects include Japan’s consistent investment in supercomputing, including the recent £585 million Fugaku, and the Australian National Research Infrastructure. There have been proposals for investment in national compute infrastructure specifically for AI such as the US NAIRR, the UK’s AIRR and exascale, and the EU’s EuroHPC (Belfield, 2023). A US-led Allied Public-Private Partnership for Frontier AI could be an international complement to, or a substitute for, these national projects. 

A US-led Allied Public-Private Partnership for Frontier AI would pool funding and resources to build the capital-intensive, expensive, specialized, joint infrastructure for frontier AI. The costs of these frontier training runs and clusters have already stretched beyond the budgets of academic and open-source groups, and has led all frontier AI companies to enter into compute partnerships with Big Tech hyperscalers. On current cost trends, start-up, academic or open-source projects are ruled out - as indeed are almost all companies, apart from Microsoft, Alphabet, Meta, Amazon; the AI companies in ‘compute partnerships’ with them; and their Chinese equivalents. 

Costs are currently \$10-100 million for a frontier training run.\footnote{In amortised costs. Total costs are closer to \$1 billion.} If historical trends continue, costs will be in the \$10-100 billion range within a decade (Cottier, 2023; Sevilla et al., 2024). The cost of building the compute cluster on which a frontier training run is conducted is typically around an order of magnitude more. GPT-4 (2e25 FLOP) was trained on 10,000x more compute than GPT-2 (\(10^{21}\) FLOP). By 2030, another 10,000x jump is possible to a \(10^{29}\) FLOP frontier training run. The data centre for this would use tens of millions of H100-equivalent GPUs and 5 gigawatts (GW) of energy - the equivalent of the largest nuclear power plants we have, the energy demand of a 1 million person city.  This would require cumulative investments of hundreds of billions of dollars. This could be within the reach of the corporate R\&D budgets of the biggest companies in the world, and yet building a cluster and conducting a training run with costs in the hundreds of billions would be one of the biggest corporate investments ever: “betting the company”.\footnote{ Some of the biggest companies in the world like TSMC or Exxon have annual capital expenditure (CapEx) budgets of around \$20-30 billion (Jewell, 2024). Some large corporate megaprojects like oil fields, LNG terminals or public transport infrastructure have cost \$20-50 billion.} There is even some speculation about a ‘trillion-dollar cluster’ drawing on 20\% of US electricity production (Aschenbrenner, 2024) which would likely be out of the reach of companies and require a state effort. The alternative to a Big Tech project or a single state project could be to pool funding to build the infrastructure for frontier training runs conducted on extremely large compute clusters. 
\begin{table}
\centering

\begin{tabular}{|>{\raggedright\arraybackslash}p{0.15\linewidth}|>{\raggedright\arraybackslash}p{0.1\linewidth}|>{\raggedright\arraybackslash}p{0.1\linewidth}|>{\raggedright\arraybackslash}p{0.1\linewidth}|>{\raggedright\arraybackslash}p{0.1\linewidth}|>{\raggedright\arraybackslash}p{0.1\linewidth}|>{\raggedright\arraybackslash}p{0.1\linewidth}|>{\raggedright\arraybackslash}p{0.1\linewidth}|}\hline

 \textbf{1 trillion +}&  &  &  &  &  &  & \textit{Cluster}\\\hline
  \textbf{100 billion – 1 trillion}&  &  &  &  &  & \textit{Cluster }& \textit{Training run}\\\hline
  \textbf{10 billion – 100 billion}&  &  &  &  & Cluster & \textit{Training run}&  \\\hline
  \textbf{1 billion – 10 billion}&  &  &  & Cluster & \textit{Training run}&  &  \\\hline
  \textbf{100 million – 1 billion}&  &  & Cluster & \textit{Training run}&  &  &  \\\hline
  \textbf{10 million – 100 million}&  & Cluster & Training run &  &  &  &  \\\hline
  \textbf{1 million – 10 million}&  & Training run &  &  &  &  &  \\\hline
  \textbf{100,000 –  1 million}& Cluster &  &  &  &  &  &  \\\hline
  \textbf{10,000 – 100,000}& Training run &  &  &  &  &  &  \\\hline

 & \textbf{GPT-2} & \textbf{GPT-3} & \textbf{GPT-4} & \textit{\textbf{GPT-5}}& \textit{\textbf{GPT-6}}& \textit{\textbf{GPT-7}}& \textit{\textbf{GPT-8}}\\\hline
 & \multicolumn{7}{|c|}{\textbf{Generation of frontier model}} \\ \hline 

\end{tabular}
\caption{\textbf{Estimates and predictions for the cost of building a frontier cluster and conducting a frontier training run across seven ‘generations’ of frontier models.} Cost range in orders of magnitude (OOMs) in 2023 dollars. \textit{Italics indicate a prediction for future clusters and training runs based on cost trends and corporate investment plans.} See Appendix for more details and sources.}
\label{table:5}

\end{table}

\subsection{Advantages and disadvantages of different projects to develop and deploy frontier AI}

A US-led Allied Public-Private Partnership for Frontier AI would be a significant undertaking which could raise serious concerns. However, the costs, benefits, and risks of such a project cannot be assessed in isolation. The relevant question is: what are the advantages and disadvantages of different Western projects to develop and deploy advanced AI? By advanced AI, I mean frontier AI systems all the way up to and including Artificial General Intelligence (AGI). If costs for building clusters and conducting frontier training runs indeed grow to tens and hundreds of billions of dollars, then start-up, academic or open-source projects are not feasible alternatives. There are only three alternatives: Big Tech companies, national governments or a US-led Allied project. Each of these three alternatives comes with their own set of crucial considerations.

What properties should we want from a project to build advanced AI systems, up to and including AGI? Let us consider seven desiderata. A project should ideally be legitimate, secure, safe, non-adversarial, peaceful, not prone to misuse, and feasible. By ‘\textit{legitimate}’, I mean a process that can be and is widely accepted as fair, appropriate or right and that broadly distributes the benefits of advanced AI. In particular, the societal changes associated with advanced AI may be so profound as to require democratic consent. By ‘\textit{secure}’, I mean with sufficient physical, personnel and cybersecurity to prevent exfiltration to users who may misuse the system. By ‘\textit{safe}’, I mean a process that does not deploy advanced AI until and unless it has been rigorously tested and demonstrated to be safe and aligned and will not lead to catastrophic accidents. By ‘\textit{non-adversarial}’, I mean a process without harmful race dynamics that could lead to shortcuts on security, perhaps even a process with some degree of coordination or cooperation between actors. By ‘\textit{peaceful}’, I mean a process that does not run the risk of excessive instability or violent conflict - for example instigated by a state that perceives the development or deployment of advanced AI as a security threat. By ‘\textit{prone to misuse}’, I mean a process that raises the risk of the AI deployer misusing AI in an unethical manner, for example to harm others’ rights and interests by manipulating or controlling them - especially AI misused for power concentration or democratic erosion within democracies.  Finally by ‘\textit{feasible}’, I mean having sufficient possibility of actually occurring - not being too difficult to accomplish.

\begin{table}
\centering

\begin{tabular}{| l  |l  |l  |l |}
\hline
 & \textbf{Private} & \textbf{National} & \textbf{US-led Allied}\\
\hline
\textbf{Legitimate} & 3\cellcolor{red} & 2 & 1\cellcolor{green} \\\hline
 \textbf{Secure}  & 3\cellcolor{red} & 1\cellcolor{green} &2 \\\hline

\textbf{Safe} & 3\cellcolor{red} & 2 & 1\cellcolor{green} \\
\hline
\textbf{Non-adversarial} & 2 & 3\cellcolor{red} & 1\cellcolor{green} \\
\hline
\textbf{Peaceful} & 2 & 3\cellcolor{red} & 1\cellcolor{green} \\
\hline
\textbf{Not prone to misuse} & 2 & 3\cellcolor{red} & 1\cellcolor{green} \\
\hline
\textbf{Feasible} & 1\cellcolor{green} & 2 & 3\cellcolor{red} \\
\hline

\end{tabular}

\caption{\textbf{Comparing seven properties we might want from a project for developing advanced AI across three alternatives: a private project, a national project and an international project.} Red and 3 is the lowest, green and 1 is the highest.}
\label{table:6}

\end{table}

There is no consensus within the field of AI governance on which actor(s), or set of relationships between actors, is more likely to achieve these goals. Corporate projects face significant legitimacy problems after a certain stage of development: a private actor taking large-scale decisions that could affect humanity as a whole, with little guarantee that the benefits will be broadly shared. Companies are less secure than states, and do not have sufficient mechanisms to guard against misuse. Moreover, companies are profoundly affected by and contribute to race dynamics: they face strong incentives from market competition, shareholders or from their finite cash reserves to quickly develop and deploy systems to the market, and do not have to internalise externalities and risks unlike a national or Allied project (and so may be less safe). However there are currently several corporate projects, so they are the most feasible.

National projects also face legitimacy questions after a certain stage: one country taking large-scale decisions that could affect humanity as a whole. They also face incentives (from perceptions of gaining national advantage) to develop and deploy systems quickly compare to a US-led Allied project with a clearer lead over its adversaries. A national project would be a significant concentration of power and could be misused to undermine democracy. Most importantly however, they also raise concerns that they will become adversarial and non-peaceful: other states may be concerned that a state with a national project would use its advanced AI against them to undermine their security, sovereignty, prosperity and prestige. These other states might respond by racing - which could lead to shortcuts on security or safety - and by threatening instability or conflict. On the other hand, a national project may be the most secure, as it can employ strict personnel, physical and cybersecurity standards. 

A US-led Allied project faces fewer legitimacy concerns, if it is formed by several democratically accountable governments. It would be less secure than a national project, but more secure than a corporate one. By consolidating frontier training runs at one cluster and including several decision-makers it would likely be safer and less prone to misuse. It would be better placed to signal to adversaries and so could also be less adversarial and more peaceful - but we might have questions about its feasibility. Obviously much more can be said for and against all these alternatives, and significantly more research and discussion should be conducted. This is a major debate, and one I seek to contribute to rather than attempt to resolve.

\subsection{Advantages and disadvantages to the US Government of including allies and companies}

Should the USA include its allies in such a project - or just go it alone?\footnote{This question has repeatedly been posed to me like so: “\textit{When} the USA does a Manhattan Project for AI, should it include its allies?"} Should it structure this project as a public-private partnership or a state-run project? There is no consensus within the field of AI governance on these questions. There are three key advantages to including allies and companies - cost-sharing, access to resources, and structural features that may make a project more likely to be successful, safer, less prone to misuse and better able to signal peaceful intent. These three advantages could contribute to a US-led Allied PPP having a clear lead over any other project, such as a Chinese state-run project. There are also two key disadvantages - joint decision-making, access and benefit-sharing and security concerns, though both of these may be mitigated to some extent.

First, \textbf{cost-sharing}. The cost of buying state-of-the-art chips and building specialised frontier clusters and associated energy infrastructure could run into the hundreds of billions if not trillions of dollars. US allies can help share this cost.\footnote{Governments are likely to borrow to fund this, rather than it being funded directly from tax receipts, with a consequent pressure to increase tax or cut other government projects. So the cost of such a project would likely go onto government balance sheets as debt. In that case, a US-led Allied project would spread this debt across the balance sheets of multiple governments. This would be less like the US Government making one truly huge bet and more like several states all making similar but smaller bets. International debt markets may prefer that as the risk is more diversified, potentially meaning a overall lower cost to raise that debt.} There could be major cost concerns about a unilateral project - sharing those costs with allies could give a US-led Allied PPP an advantage over its competitors. A public-private partnership can further share this cost with the private sector. States were willing to borrow extensively for WW1 and WW2, and to a lesser extent for the Global Financial Crisis and COVID-19. By contrast, successive US Governments have repeatedly stated that governments will not borrow trillions to pay for the clean energy transition. Instead, they have sought to derisk private investment and leverage limited government funding into multiples more of private funding. Advanced AI seems less likely to be viewed as an immediate security or financial crisis and more a long-term challenge and opportunity like climate. More importantly, companies and investors see a clear financial return from frontier AI and are already willing to invest tens of billions - rather than cut them out, the USG would likely prefer to roll that investment into a larger project. So there are cost-sharing advantages to a US-led Allied public-private partnership that can leverage funding from allies, Big Tech companies and large institutional investors.

Second, \textbf{access to resources}. The resources needed to conduct frontier training runs include chips, talent, data and energy. Almost all of this is currently held by a few large companies, many of which are located in US allies. The chip supply chain includes US firms like NVIDIA at the design stage and US hyperscalers at the cloud stage, but in between the supply chain is dominated by companies in US allies: ASML in the Netherlands, companies in South Korea and Japan, and most importantly TSMC in Taiwan. The US is trying to bring some of that supply chain on-shore with the CHIPS Act and massive subsidies to TSMC, but it is unclear to what extent that will be effective. China has been unsuccessfully trying to indigenise a supply chain for decades; TSMC plans to keep its leading node capacity in Taiwan. These compute producer allies can contribute compute infrastructure. The UK and Canada can contribute their AI talent - as well as their deep trusted relationships of intelligence-sharing with the USA. Some countries such as Canada, France and Scandanavia have both low cost of industrial energy and substantial energy generation capacity that can be used to power data centres. Allies have resources that the USA needs, and can contribute these resources to a US-led project. This could give an advantage to a US-led Allied PPP over rival projects.

Additionally, the USA’s need to access these resources gives allies leverage. These allies’ preference would be to be involved in a US-led project - paying their way, sharing costs, and providing resources to a US-led project. However, if the US were unrelenting on excluding its own allies and going it alone, these allies might have a credible alternative: a joint project without the US. The European Commission is reportedly considering proposals for €100 billion over the next five to seven years in EU AI research, explicitly inspired by CERN (Wold, 2024). This could form the nucleus of a joint project. It could combine the chip supply chain of the Netherlands and East Asia, AI talent from the UK and Canada, and the cheap and abundant energy of Canada, France and Scandinavia. Withholding these resources from a US-only project and instead allocating them to a joint project could make that joint project competitive with a US-only project, and would in any case slow down a US-only project in comparison to China. The US should want to avoid those situations by having a clear lead, so it has an interest in including allies. A joint project would face challenges that undermine its credibility. For example, East Asian allies rely on the USA for their defence against China. If the USA threatened to withdraw that support, would they provide for their own defence, or might the UK or France expand their umbrellas? Both options seem doubtful. Overall, this is very much a second-best alternative: allies' strong preference would be to support a US-led project. However, its potential credibility might give these allies some leverage to encourage the US to include them properly. 

Tapping companies' expertise and resources is another reason for a public-private partnership. Nationalisation or a fully state-run ‘Manhattan Project' equivalent seems very unlikely due to cost, legal challenges, and lack of state capacity, ideology and practice. Nationalisation would either involve paying market rate for these companies, with costs in the trillions, or expropriation, which would be strongly legally and politically opposed. A state-run project could also exclude companies from operating at the frontier, and would be opposed by those companies.\footnote{Moreover, a significant portion of the value of US Big Tech companies is tied up with their AI prospects. At an individual level, much of the compensation at AI companies is in the form of equity, so executives' wealth is dependent on being able to capture value from frontier AI. At a macro level, the tech sector accounts for around a third of the value of the entire S\&P 500. A bet on the US stock market is largely a bet on Big Tech, and thus a bet on AI. If companies are not able to capture value from frontier AI, this could damage the entire stock market. No US Government would want to tank the stock market like this.} In general, large nationalisations and large state-run projects are things of the past - the US government has had little interest or experience in them for decades. A public-private partnership with US companies seems more likely. In such a situation, the US government would want the active and willing partnership of US companies, not begrudging participation. This may also be important legally, to stave off possible lengthy lawsuits. The main legal authority the US Government might rely on, the Defence Production Act, is structured around requiring companies to complete government orders first before any other orders, rather than being more directive or intrusive. An explicit public-private partnership in which these companies are able to profit from frontier AI is the most likely way to have active, willing participation from these companies.

Third, \textbf{structural advantages}. US elites and the US public should and would be concerned about any project that is unsuccessful, unsafe, prone to misuse or could spark international conflict. Pursuing a US-led Allied public-private partnership rather than a unilateral project could be more successful, safer, less prone to misuse, and less strategically destabilising, in ways that benefit US national security. As an additional benefit, an adversary may be less able to present a joint project as an illegitimate unilateral project.

Most clearly, a US-led Allied public-private partnership would have a clear lead over any alternative, as these countries are the main or sole producers of state-of-the-art chips and have a joint clear lead in talent. This ensures that the USA (and its allies) will be the first to gain economic and strategic advantage from this powerful new technology. As an additional benefit, this lead would reduce pressures to race to beat a competitor, meaning the project can invest the proper time and resources into security and safety - to ensure that the project is secure, and the AI systems have been rigorously tested and demonstrated to be safe and aligned and will not lead to catastrophic accidents. 

An US-led Allied public-private partnership also has some clear advantages when it comes to transparency, debate and decision-making compared to a unilateral project. These advantages could make the project safer and less prone to misuse, thus better for US national security. At the extreme of how closed off a national project can be, let us consider the Manhattan Project. The US federal Government borrowed money for the war effort using bonds; put a significant part of that state funding into a secret trust account that was kept secret from the Cabinet, Vice President and Congress; and gave the head of the Project Leslie Groves a “AAA priority" to request critical materials in a near command-and-control war economy (Rhodes, 1986). There was no public debate on the race for the bomb, its testing or its use, and no decision-making or access was given to allies.\footnote{ The Manhattan Project had some limited participation from the UK and Canada, but this mainly consisted of them handing over research and some scientists to a US-only project. The UK and Canada were excluded from decision-making and denied access to the Bomb after the war.} A US-led Allied public-private partnership would be very different - there would be extensive public debate in several countries over such a large and expensive program; allies would want some insight and consultation and companies and investors would want clear terms of access and benefit-sharing - otherwise allies and companies would not invest and participate. It is unlikely that decisions would be made by voting or consensus - more likely is that allies and companies would be notified and consulted, with decisions ultimately being made by the US Government. However, a US-led Allied public-private partnership would have to act in a manner that most participating states were more or less on board with - otherwise they could withdraw participation. Such a structure may result in better decisions than a secret, unilateral structure. A public-private partnership involving allies means that a process of generating, sharing and discussing evidence has to happen - rather than decisions being (perhaps with little evidence behind them) at the top of a hierarchy and subordinates following orders. This creates better incentives for progressing safely and securely and reduces the chance of a single unilateral decision-maker taking a very risky decision. This can also help address the concern that advanced AI capabilities may be misused within democracies to undermine or erode democracy. 

A US-led Allied public-private partnership may be better able to signal peaceful intent to adversaries, reducing the chance of escalation, miscalculation and conflict. Adversarial states are likely to be particularly concerned that a US-only project is not peaceful – that the US Government would ‘misuse’ (from its perspective) advanced AI to undermine their security, sovereignty or regime stability. Assuming that they could not win any ‘race’ to a given level of capability, such an adversary has only three options: acquiesce to US hegemony, reach an agreement with the USA, or threaten escalating coercive action (Belfield, 2022). Escalatory attempts at coercion threaten US national security, as they could include sanctions, espionage, cyberattacks, blockades or even threats of kinetic strikes. Including allies could reduce an adversary’s concern and so reduce the threat to US national security. For example, one highly destabilising factor in international relations is fear of a surprise attack. An adversary could fear that one national leader could unilaterally and secretly decide to take a surprise first move against it using advanced AI. An adversary would be less afraid of surprise attack from a joint project, as it would be hard to secretly persuade allies and companies to sign on to such a plan. So a US-led Allied PPP could reduce an adversary's concern about a surprise attack. A joint project also makes agreements with adversaries more feasible, as specific technical processes would likely be set up within the project that would generate evidence that could also reassure adversaries. For example, a national project could better hide that a frontier training run had started - this would be harder for a joint project. A joint project would have to share some information internally on when a frontier training run is going to occur, how long for, using how much compute and using what datasets. Some of that information could be shared with adversaries, in ways that do not risk espionage and sabotage. This could reassure an adversary without giving away an advantage to it. For example, a joint project might generate evidence in the form of pre-training run risk assessments and pre-deployment evaluations. The redacted results of these could be shared with an adversary to reassure them and dissuade them from preventative escalation, without conceding any advantage.

The three advantages are cost-sharing, access to resources, and structural advantages. However, there are also two key disadvantages to the US Government of including companies and allies - concerns around joint decision-making, access and benefit-sharing and security concerns.

Fourth, \textbf{joint decision-making, access and benefit-sharing}. Governance of a joint project could take different forms, drawing on different international institutions. One straight-forward approach would be consensus-based, wherein all participating states would have to agree on any decisions. This is the model of the UN Security Council. Another model involves simple majority voting, with each participating state casting one vote. This model mirrors the structure of the IAEA's General Conference. A third option could emulate the decision-making processes of the International Monetary Fund and the World Bank, where voting rights are initially allocated based on a one-state-one-vote policy and then supplemented according to different funding contributions. However, all these options risk gridlock, or the US Government being outvoted. More likely then is a US-led Allied public-private partnership in which the US Government is the primary decision-maker, and other participants are notified or consulted, but do not have explicit decision-making rights within the project. However, there could still be scope for ‘voice' and ‘exit' (Hirschman, 1970): allies and companies would likely reach agreements before joining such a project, would be able to raise concerns and preferences throughout and could always withdraw participation. The USA would have the greatest access to advanced AI, and gain the most benefits from it - but the US Government may simply not want to share any access to or any of the benefits of advanced AI with any other state or with companies. Such maximalism is difficult to reconcile with a joint approach, and could perhaps only be achieved through a more expensive and difficult unilateral project. The Joint Strike Fighter program could be a useful analogy - allies have limited access while the US Government has first access to its advanced capabilities, and gains the greatest defence benefit overall.

Fifth and finally, \textbf{security}. A US-led Allied public-private partnership could develop advanced AI capabilities that might reshape the international distribution of power. The USA would not want those capabilities, the AI systems that enable them, or technical insights into building those AI systems to be stolen or sabotaged by adversaries. The US Government may worry that including allies into a PPP could increase the risk of successful espionage or sabotage by adversaries.\footnote{The figure that haunts these concerns is Klaus Fuchs. Despite being a German (who became a refugee after the Nazis came to power), and despite being only one of the many ‘atomic spies' for the Soviet Union - including several Americans such as the Rosenbergs, the Greenglasses and the Cohens - Fuchs was perhaps the most important Manhattan Project spy, and he had British citizenship.}

These are legitimate and serious concerns, but they may be able to be addressed and mitigated to some extent. The USA participates in several deep, long-standing, highly secure partnerships with its allies. It shares highly sensitive military intelligence with its NATO or AUKUS allies - and develops technology in partnership with them. It shares even more highly sensitive intelligence and access with its Five Eyes friends. In each of these cases, US allies have been able to provide reassurance to the USA about their level of security sufficient to enable a deep trusted relationship. The US-led Allied Public-Private Partnership for Frontier AI could be structured similarly, requiring very strict information security. For example, its AI clusters would be located within the USA, and the US national security community would run its physical, cyber and personnel security.

In ‘Securing Model Weights’, Nevo et al propose an influential set of five Security Levels: SL1 to SL5 (Nevo et al., 2024). RAND L4 and RAND L5 are necessary to protect against determined state adversaries. Companies are currently not meeting SL3, so cannot “likely thwart cybercrime syndicates or insider threats”, let alone “operations by leading cyber-capable institutions” like states. In time, companies may need to seek advice and support from their state’s intelligence and security agencies - especially at the highest end of SL5 to be secure against a state actor. 

Full implementation of SL4 or SL5 would require some wrenching cultural changes. For example, take personnel security and a robust insider threat program. At many government facilities an employee needs to have citizenship or even be born in the country, and to not have close friends or family in adversarial states who could be threatened in order to blackmail that employee. Security clearance can involve intrusive vetting - sharing details about finances, sexual preferences and drug habits - and restrictions on travel, such as pre-approval for going abroad to scientific conferences. Or take physical security. At many government facilities staff work inside a Sensitive Compartmented Information Facility (SCIF) - a highly secure enclosed area. Working in a SCIF is like working in shipping container: SCIFs do not have windows, they are fairly small and cramped to maximise efficiency, and though there is air-conditioning they can become hot and stuffy. Finally, take cybersecurity. At many government facilities, staff are not allowed to take personal devices into the building, and consumer-grade productivity and collaboration tools like Slack or Zoom may not be allowed. These levels of security requirements would be a big change from current cultural norms in AI companies, and are likely to severely impact recruitment and retention, and the speed of R\&D. Some of them might be discriminatory if done by a private company without government mandating them to do so. For example, in the USA, one cannot discriminate against non-citizens unless required to do so by the US Government. For these reasons, companies are unlikely to move to SL4 or SL5 unilaterally and voluntarily, unless they and their competitors are all required and assisted to do so by the Government.

Note too that securitisation can have significant additional downsides. Secrecy can lead to a lack of debate and sense-checking about facts or policy options. For example, secrecy, securitisation and compartmentalisation was a key contributing factor to various Cold War panics that encouraged arms racing: from the US “bomber gap" and “missile gap" to the Soviet fear of falling behind on biological weapons (Belfield \& Ruhl, 2022). Possible mitigations include: encouraging robust internal policies and culture of debate (like 1950s RAND), strong internal whistleblower protections, and robust Congressional and judicial oversight.

A US-led allied public-private partnership has certain advantages to the US national interest. It shares high costs, enables access to scarce resources, and has certain structural advantages that could make it more successful, safer, less prone to misuse, and better able to signal to adversaries. However, questions of decision-making and security would need to be satisfactorily addressed.

\subsection{Next steps to a US-led Allied Public-Private Partnership for Frontier AI}

A stylized path to a US-led Allied Public-Private Partnership for Frontier AI could proceed as follows. 

\begin{table}
\centering

\begin{tabular}{|>{\raggedright\arraybackslash}p{0.18\linewidth}|>{\raggedright\arraybackslash}p{0.18\linewidth}|>{\raggedright\arraybackslash}p{0.18\linewidth}|>{\raggedright\arraybackslash}p{0.18\linewidth}|>{\raggedright\arraybackslash}p{0.18\linewidth}|}
\hline
\multicolumn{5}{|c|}{US-led Allied Public-Private Partnership for Frontier AI} \\
\hline
\textbf{Phase 1} & \textbf{Phase 2} & \textbf{Phase 3} & \textbf{Phase 4} & \textbf{Phase 5} \\
\hline
\textbf{Agreement on pilot project} & \textbf{Pilot project} & \textbf{Incrementally increase investment} & \textbf{Fold in remaining projects} & \textbf{Agreements with adversaries}\\
\hline
The structure, funding, private partners and locations of the pilot project are agreed between participating states. The Public-Private Partnership (PPP) uses advanced market commitments to secure the next generation of state-of-the-art AI chips.& The PPP uses the jointly acquired state-of-the-art AI chips to build a frontier compute cluster and conduct a frontier training run. This is could be a model of best practice: high-quality risk assessment and evaluations throughout. Access is provided to participating states and partners.& If the pilot project is deemed successful, the next stage would be to acquire the next generation of state-of-the-art AI chips, build the next generation cluster and conduct a frontier training run. Access is expanded, in a secure manner. & If the PPP is going well, then remaining private and national frontier projects could be folded in. State-of-the-art AI chips could be preferentially (or eventually exclusively) sold to the PPP for a specified period.& The PPP could be the primary site of arms control style monitoring and verification.\\
\hline

\end{tabular}
\caption{\textbf{The aims and stages of a US-led Allied Public-Private Partnership for Frontier AI.} It would share the rising computational costs of AI research and by centralizing the riskiest training runs in a US-led Allied PPP, it could be more succesful, safer, more secure, and reduce harmful competitive pressures.}
\label{table:7}
\end{table}

\subsection{Phase 1: Agreement on pilot project}

The first step would be for states to propose the pilot project, get buy-in and agreement, and commit capital. The relevant states that would jointly invest in an US-led Allied Public-Private Partnership for Frontier AI would be the USA and it liberal democratic allies: the UK; some EU/NATO allies like the Netherlands or France; Japan, South Korea and Taiwan in East Asia; and Five Eyes allies like Canada and Australia. These would likely be the same allied countries that might pursue the Secure Chips Agreement in the previous section, indeed participation in the Secure Chips Agreement might be a condition for participation in the US-led Allied project. 

Planning the pilot project would require navigating large capital commitments for the compute cluster. During this planning Phase, the leading generation of AI chips would already be commissioned and ordered, so it would likely not be viable for the US-led Allied PPP to compete to acquire \textit{this} generation of advanced AI chips. Instead, the project could agree to purchase the \textit{next} generation of leading AI chips to build a cluster for training state-of-the-art AI model as an advance market commitment (Kremer, Levin and Snyder, 2020). This approach might also be more in keeping with the broadly slower procurement processes in the public sector relative to the private sector.

The US-led Allied project would likely develop its cluster as a public–private partnership. One analogy is the Joint Strike Fighter program, which has one lead private contractor and multiple private sub-contractors from different participating countries. The PPP would partner with private sector expertise like leading chip designers and fabs; hyperscaler cloud providers with experience operating these advanced compute clusters; and with leading frontier AI model developers. The pilot project would not seek to compete with these companies, but to partner with them. The key contribution of the US Government would be to broker this relationship; ensure the PPP can quickly and reliably secure land, permits and energy; and ensure high levels of cybersecurity, personnel security and physical security.

One question would be the location of the infrastructure of a US-led Allied PPP. A US-led Allied PPP could be a distributed, federated, ‘virtual’ cluster operating across multiple data centres, perhaps even in multiple countries. This could, for example, leverage existing proposals such as the US NAIRR or EU EuroHPC. However, this might have several drawbacks, including lower efficiency and higher cost\footnote{ For example, in one experiment a small 1.3 billion parameter model was 1.7-2.3x slower to train in a federated manner (Yuan et al; 2022).}; worse information security and harder monitoring/verification; and less ability to clearly signal that a ‘secret’ training run is not being pursued. The current training paradigm is to use a geographically close number of data centres. The USA has the energy capacity, AI companies, and security and would lead a US-led Allied PPP coalition. Several factors might influence the location decision within the USA. A nearby international city could enable scientific collaborations - like CERN and Geneva (Fischer and Wenger, 2019). Cost, availability, and reliability of electricity and land, and stability (geological stability or lack of extreme weather events) - could all be important for locating the physical data centres. There is also always a political factor in where important government expenditures are made within the USA.

\subsection{Phase 2: Pilot project}

The pilot phase would use the jointly acquired world-leading compute cluster to train a frontier foundation model. The primary goal of this training run would be to demonstrate feasibility of the project and produce a frontier model. However, it could also be used as a model for best practices and a demonstration of possibly stringent standards: high-quality risk assessments, and robust testing and evaluations before, during and after the training run. An important joint benefit for states - and a requirement of the project - would be the robust information security of the cluster, sufficient to be resilient to state-level espionage.

Several strategies can be considered for disseminating access to this trained model to users. One method might be structured access, where users in all participating nations could obtain API access centrally from the US-led Allied PPP. This has the advantage of information security and lower risk of exfiltration or misuse. However, participating states and companies may want access to the trained model weights itself - and there could be efficiency and latency benefits of conducting inference at several locations in particpating states. Alternatively, with a sufficiently secure information system, the model weights could be securely transferred to licensed entities in each participating country, whether they be private corporations or public agencies. These licensed entities could in turn fine-tune the model for particular use-cases and offer API access within their countries.

This scenario is analogous to a central uranium enrichment plant that produces low enriched uranium (LEU), which is then securely transported to nuclear power stations.\footnote{ This shares some similarities with Eisenhower’s initial suggestion to the UN General Assembly in September 1954: an international ‘nuclear bank’ with control and custody of fissile material. This proposal eventually turned into the IAEA instead.} In the same way that nuclear power plants need to be monitored to prevent the diversion of LEU and plutonium, the licensed entities would need to be monitored to prevent copying/transfer of the trained model, and to prevent some kinds of fine-tuning that would give the model dangerous capabilities. For uranium, this process occurs under the watchful eye of the IAEA. Similarly, the IAIA from a previous section could oversee the US-led Allied PPP, the transfer process, and the licensed entities.\footnote{ What about access and benefit sharing beyond the participating states, for example for the Global South? One line of argument for global access and benefit sharing is moral. All should benefit from such a powerful new technology with the potential to benefit everyone. There is a fairness argument - all people in the world share some level of risk from AI development so also should share the benefits. However many international agreements look more like ‘deals’. In the Convention on Biological Diversity (CBD), the Nagoya Protocol on Access to Benefit Sharing or Cartagena Protocol on Biosafety can be seen as a ‘benefit-sharing in exchange for Global South genetic resources or biosafety’ deal. Genetic resources may be analogous to the ‘data commons’ used for pre-training (Chan, Bradley and Rajkumar, 2023). Just like biosafety failures in the Global South threaten rich countries, so too might AI misuse in the Global South (Abungu, Malonza and Adan, 2023). Exploring what bargaining power the Global South might have and exploring the nature of such potential ‘deals’ requires further research.}

\begin{figure}
    \centering
    \includegraphics[width=1\linewidth]{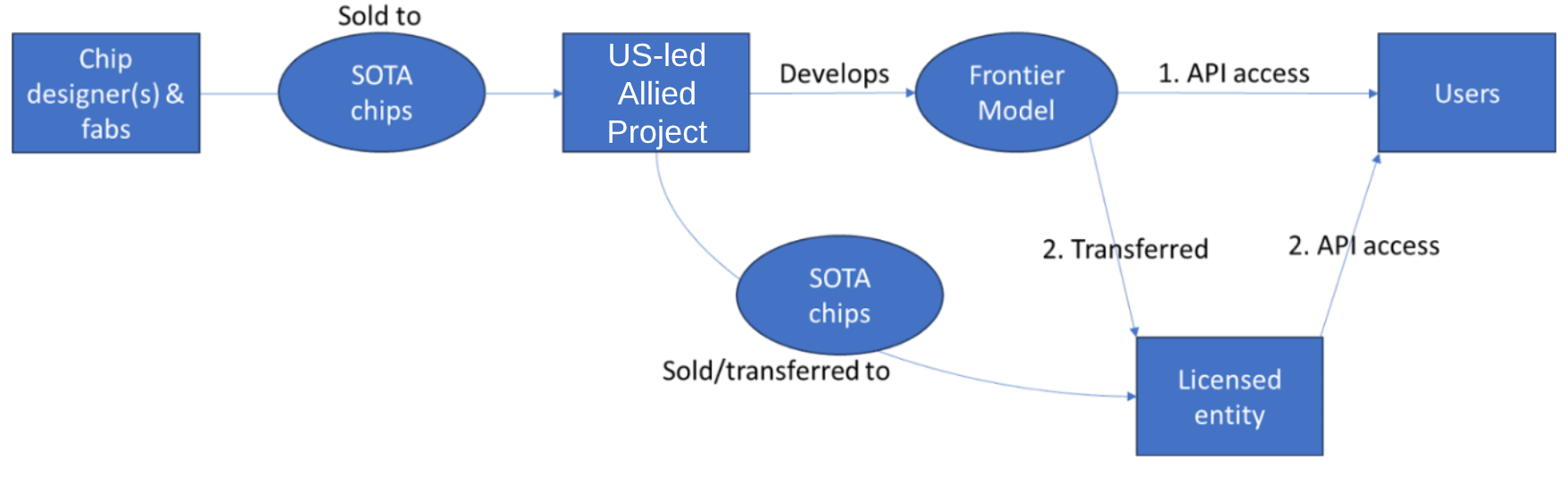}
    \caption{\textbf{Figure 12: The flow of state-of-the-art chips and frontier models in a US-led Allied PPP for frontier AI.} A US-led Allied PPP could purchase the state-of-the-art chips from the chip designers and fabs - perhaps exclusively. Once the state-of-the-art chips are used to train a frontier model, they can then be sold or transferred to licensed entities (public or private) in participating states. Access to the trained model could occur in two ways. Option 1 involves direct API access by users to the frontier model, Option 2 involves the transfer of the frontier model to a licensed entity in a participating state, which would then offer API access to users.}
    \label{fig:7}
\end{figure}

\subsection{Phase 3: Gradual Scaling of Investment Informed by Pilot Project}

If the pilot project is successful, the next stage would be to acquire the next generation of advanced AI chips, build the next generation supercomputing cluster, and train the next generation foundation model.

To return to the imperfect CERN analogy, one of the merits of CERN is its modular structure which allows for the gradual escalation of investment with each new particle accelerator. The Large Hadron Collider (LHC), for example, is but the latest of several particle accelerators and colliders at CERN which have gradually grown more expensive, bigger, and more powerful. The decision to fund and build each new machine allowed CERN to grow incrementally. 

Analogously, a US-led Allied PPP for frontier AI could incrementally and methodically boost investment, provided the pilot project and successive stages are successful. The decision to fund and build each new supercomputing cluster and to conduct each frontier training run would each be an incremental step in building a US-led Allied PPP for frontier AI. On current trends this decision to go to the next stage will happen roughly every one to two years. There are several benefits to this approach. It avoids the participating states and companies having to make a huge initial investment - instead it can be ramped up over time. They can experiment without having to commit fully immediately, instead building up confidence and trust over time. 

\subsection{Phase 4: Fold in remaining projects}

Should a US-led Allied PPP for frontier AI progress satisfactorily, private enterprises and national foundation model developers could be fully integrated into the project. This would lead to a situation where most (or even at some point all) the largest - and therefore riskiest - frontier training runs were being conducted at the Allied Project for frontier AI. The process by which this could occur would not be 'banning' private developers from conducting frontier training runs. Instead it would occur through the allocation of state-of-the-art chips. At a hardware level, the most advanced chips, optimized for large training runs, could be preferentially (or even at some point exclusively) sold to the US-led Allied PPP for frontier AI. Once the state-of-the-art chips are used to train a frontier model, they can then be sold or transferred to licensed entities in participating states. The US-led Allied PPP by that point would have moved on to securing the next generation of state-of-the-art chips, so the previous generation could be sold or transferred. Instead of independently developing their own frontier models, private developers would be part of the PPP. If not, they could develop their own foundation models (smaller than frontier-scale) or build upon the models generated by the Allied Project for frontier AI, like today downstream developers build upon models created by upstream foundation model developers.

Such centralisation would be a major intervention into the market. I discussed both the positives and negatives of this extensively above. Briefly, a US-led Allied Public-Private Partnership would have a clear lead over all rival projects - with the benefits of securing economic and security advantage for the USA and its allies, as well as reducing 'arms race' incentives and incentives to cut corners on security and safety. It would also allow for much higher information security to prevent state espionage, higher legitimacy, and higher security and safety standards compared to private companies. However, such an intervention would be a concentration of power. It should be noted that if cost trends for clusters and frontier models continue to scale to hundreds of billions of dollars, then it is \textit{this} trend by itself that will concentrate the ability to conduct frontier training into a tiny number of hands. Already in 2024 only the four or five biggest Big Tech giants (and those in compute partnerships with them) are able to operate on the frontier. Power is already heavily concentrated. Nevertheless, shifting to a situation where most advanced chips were sold to a US-led Allied PPP for frontier AI first, and most frontier training runs were conducted by that public-private partnership first, would be an additional concentration: from perhaps four projects to perhaps one. This has certain downside risks, such as on the margin limiting these few individual companies’ ability to innovate, experiment and develop AI as they see fit; and stifling the diversity of approaches that may emerge from more market-driven competition by instead encouraging a one-size-fits-all approach to pre-training that may have unintended consequences. A US-led Allied PPP for frontier AI would be more bureaucratic than pure market development. Moreover, it would need to have strong mechanisms of scrutiny and accountability such as Congressional and judicial oversight to prevent misuse. As discussed above, any possible downsides need to be both weighed alongside the possible gains to legitimacy, security, safety, non-adversariality, peacefulness and potential for misuse, and compared to the two alternatives (a single corporate or national project). This paper does not aspire to settle this debate, but rather to explore it.

\subsection{Phase 5: Agreements with adversaries}

Inviting adversaries to participate in the project - folding in \textit{their} projects to a a US-led Allied PPP for frontier AI too - seems unlikely. The PPP would develop advanced foundation models that may be strategically important and carry significant safety and security risks. Including an adversary would pose an unacceptably high risk of espionage and theft.

A US-led Allied PPP, like the Secure Chips Agreement, would continue and deepen the current situation in which US adversaries - most importantly China - are kept from competing at the cutting edge of frontier AI by restrictions on their ability to buy or build state-of-the-art AI chips. A US-led Allied PPP would extend that situation. Adversaries would be likely to critique it, as they critique the export controls. Adversaries could feel threatened if the development of advanced AI had the potential to lead to capabilities that could threaten their security, sovereignty and regime stability.

However, a US-led Allied PPP for frontier AI could play a key role in managing tensions with adversaries. For example, it would provide reassurance to adversaries that there were no secret clusters and training runs that the adversary did not know about and should be concerned about. This is because it would be structured around the incremental ratcheting up of investment: the larger the scale of the investments, the less likely it is that one of the participants could be hiding a similarly sized secret project. 

Most clearly, a US-led Allied PPP for frontier AI could become the primary site of arms control-style monitoring and verification for frontier AI.  Other states could be reassured that such a project does not threaten them through credible and verifiable commitments. It could be the organization that most often shares information, similar to information-sharing provisions in arms control agreements, and could even be a primary site for inspections. This would require very high information security. However, after decades of arms control negotiations the main states now have deep expertise in monitoring and verification in adversarial settings. This can be leveraged for frontier AI, drawing on helpful features of frontier AI such as the ability to do cryptographic, zero-knowledge, remote monitoring. 

The low-trust environment between competitor nations is reflected in President Reagan’s favourite Russian proverb: “trust, but verify.” Just as the START treaties between Russia and the United States require intensive bilateral monitoring with custom privacy-preserving hardware, software, and third-party auditing, a low-trust regime for international monitoring may have to rely on similar mechanisms to provide assurances to all parties. Though this beyond the scope of this paper, perhaps this could be a fifth scenario for an international AI governance institution: a new START for AI.

\pagebreak

\section{Conclusion}

This paper has explored four different scenarios of international institutions for the governance of AI. Together these four steps - domestic frontier AI regulation, an International AI Agency, a Secure Chips Agreement and a US-led Allied Public-Private Partnership for frontier AI - could provide governments assurance that frontier AI is being developed and deployed in ways that limit the threat to their national security.

As I noted repeatedly, this analysis seeks to contribute to and expand the ongoing discussion on these technically and politically complex issues. Over the years and decades to come, there will be extensive analysis and debate. Domestic regulation of nuclear power has been developed incrementally over the past seventy years. At an international level, the IAEA is 67 years old and the NPT is 56. The international community is not going to immediately leap to a world with strong domestic frontier regulation, clear international safeguards and monitoring, a robust AI chip tracking regime, and a US-led Allied public-private joint project to conduct frontier training runs. These ideas will be developed, debated and decided over the years and decades to come. 

Nevertheless, it can be helpful to have a sense of a bigger picture. How can we get to a situation in which governments can be reassured that frontier AI is developed and deployed in a legitimate, safe, secure manner with misuse minimised and benefits widely shared? There are incremental, sensible steps that can be taken now that build on what the international community is already doing. The steps already taken will create motivations and incentives for further deepening and harmonising collaboration.

We have some sense of the next steps to take over 2025. These next steps are sensible and reasonable on their own merits. However, they can also be viewed as next steps towards a set of intersecting and reinforcing regimes and institutions that could give governments assurance that frontier AI is being developed and deployed in ways that do not threaten their national security, but instead - in the words of the IAEA Charter - contribute to “peace, health and prosperity throughout the world”.

\section{Acknowledgments}

I thank the Future of Life Insitute for the grant that made this paper possible. I thank anonymous readers, and my colleagues at the Leverhulme Centre for the Future of Intelligence and the Centre for the Study of Existential Risk, for feedback on this paper. This paper grew out of a previous paper “Computing Power and the Governance of Artificial Intelligence”. The author would like to thank the other co-authors of that paper - especially Lennart Heim and Markus Anderljung - and those who gave feedback on that paper as well.

\pagebreak

\section*{Appendix: Estimates and predictions for frontier AI costs}

This appendix gives more detail on my estimates for Table 4 in the ‘A US-led Allied Public-Private Partnership for Frontier AI’ section.

\begin{table}
\centering

\begin{tabular}{|>{\raggedright\arraybackslash}p{0.14\linewidth}|>{\raggedright\arraybackslash}p{0.1\linewidth}|>{\raggedright\arraybackslash}p{0.1\linewidth}|>{\raggedright\arraybackslash}p{0.1\linewidth}|>{\raggedright\arraybackslash}p{0.11\linewidth}|>{\raggedright\arraybackslash}p{0.11\linewidth}|>{\raggedright\arraybackslash}p{0.11\linewidth}|>{\raggedright\arraybackslash}p{0.11\linewidth}|}\hline

 \textbf{1 trillion+}&  &  &  &  &  &  & \textit{Cluster 100m H100s, 100 GW}\\\hline
  \textbf{100 billion – 1 trillion}&  &  &  &  &  & \textit{Cluster 10-30m H100s, 3-8 GW}& \textit{Training run 9e29 FLOP 2030+}\\\hline
  \textbf{10 billion – 100 billion}&  &  &  &  & \textit{Cluster 1m H100s, 1 GW}& \textit{Training run 2e29 FLOP 2030?}&  \\\hline
  \textbf{1 billion – 10 billion}&  &  &  & Cluster 100k H100s, 100 MW& \textit{Training run 1e27-28 2028?}&  &  \\\hline
  \textbf{100 million – 1 billion}&  &  & Cluster 10k H100s, 10 MW& \textit{Training run 1e26-27 2025?}&  &  &  \\\hline
  \textbf{10 million – 100 million}&  & Cluster 1k H100s, 1 MW& Training run 2e25 FLOP 2022&  &  &  &  \\\hline
  \textbf{1 million – 10 million}&  & Training run 3e23 FLOP 2020&  &  &  &  &  \\\hline
  \textbf{100,000 –
1 million}& Cluster 100s H100s, 10 kW&  &  &  &  &  &  \\\hline
  \textbf{10,000 – 100,000}& Training run 1.9e21 FLOP 2019&  &  &  &  &  &  \\\hline

 & \textbf{GPT-2} & \textbf{GPT-3} & \textbf{GPT-4} & \textit{\textbf{GPT-5}}& \textit{\textbf{GPT-6}}& \textit{\textbf{GPT-7}}& \textit{\textbf{GPT-8}}\\\hline
 & \multicolumn{7}{|c|}{\textbf{Generation of frontier model}} \\ \hline 

\end{tabular}
\caption{\textbf{Estimates and predictions for the cost of building a frontier cluster and conducting a frontier training run across seven ‘generations’ of frontier models.} Cost ranges are in orders of magnitude (OOMs) in 2023 dollars. Italics indicate a prediction for future clusters and training runs based on cost trends and corporate investment plans.}
\label{table:10}

\end{table}

Cluster sizes are in H100-equivalents. Estimates for GPT-2 (\$50,000 training run), GPT-3 (\$2.1m training run) and GPT-4 (\$41m training run) are from Epoch’s database (Epoch AI, 2024). I assume that each H100 has a maximum thermal design power (TDP) of up to 700W. I assume that 100 H100s is roughly equivalent to a cluster of 790 V100s and that GPT-3 was trained on around 1,000 V100s. Llama-3 (which is from the same generation as GPT-4) was trained on 4e25 FLOP at a 16,000 H100s 27MW data centre. The estimate for the 'GPT-5' and 'GPT-6' clusters is from public reporting on the \$100 billion, 10-30m H100s, 3-8 GW ‘Stargate cluster’ (Gardizy and Efrati, 2024). My estimates and predictions are consistent with those in ‘Situational Awareness’ (Aschenbrenner, 2024). They are also consistent with the Anthropic CEO’s investment plans: “Claude 3 cost somewhere between \$30 million and \$300 million to train, Amodei says, declining to be more specific. He predicts training frontier models in 2024 will cost on the order of \$1 billion; the trend suggests the generation after that would cost more like \$10 billion.” (Perrigo, 2024)

\pagebreak
\section*{References}

Aarne, O., Fist, T. and Withers, C. (2024) Secure, Governable Chips: Using On-Chip Mechanisms to Manage National Security Risks from AI \& Advanced Computing. CNAS. Available at: https://www.cnas.org/publications/reports/secure-governable-chips.

Abungu, C., Malonza, M. and Adan, S.N. (2023) ‘Can apparent bystanders distinctively shape an outcome? Global south countries and global catastrophic risk-focused governance of artificial intelligence’. arXiv. 

Afina, Y. and Lewis, P. (2023) ‘The nuclear governance model won’t work for AI’, Chatham House – International Affairs Think Tank, 28 June. Available at: https://www.chathamhouse.org/2023/06/nuclear-governance-model-wont-work-ai (Accessed: 17 April 2024).

Anderljung, M. et al. (2023) ‘Frontier AI Regulation: Managing Emerging Risks to Public Safety’. arXiv. 

Aschenbrenner, L. (2024) Situational Awareness: The Decade Ahead. San Francisco, CA: Situational Awareness. 

Avin, S. et al. (2021) ‘Filling gaps in trustworthy development of AI’, Science, 374(6573), pp. 1327–1329. 

Baker, J.N. (2024) ‘International law and advanced AI: exploring the levers for “hard” control’, Institute for Law \& AI, 20 July. Available at: https://law-ai.org/international-law-and-advanced-ai/ (Accessed: 12 August 2024).

Barish, B.C. (2013) ‘Subatomic Politics: Particle Physics and Science Diplomacy’, World Politics Review, 19 March. 

Belfield, H. \& Ruhl, C. (2022) ‘Why policy makers should beware claims of new ‘arms races’, Bulletin of the Atomic Scientists, 14 July. 

Belfield, H. (2022) ‘The Rival AI Deployment Problem: a Pre-deployment Agreement as the least-bad response’, EA Forum, 23 September. 

Belfield, H. (2023) Great British Cloud and BritGPT: the UK’s AI Industrial Strategy Must Play to Our Strengths. Labour for the Long Term. Available at: https://www.labourlongterm.org/briefings/great-british-cloud-and-britgpt-the-uks-ai-industrial-strategy-must-play-to-our-strengths (Accessed: 29 May 2024).

Bengio, Y. et al. (2024) International Scientific Report on the Safety of Advanced AI - Interim Report. UK \& Korean Governments. Available at: https://www.gov.uk/government/publications/international-scientific-report-on-the-safety-of-advanced-ai.

Brass, A. (2024) Location Verification for AI Chips. Asher Brass. Available at: https://www.iaps.ai/research/location-verification-for-ai-chips (Accessed: 8 August 2024).
Brundage, M. et al. (2018) ‘The Malicious Use of Artificial Intelligence: Forecasting, Prevention, and Mitigation’, arXiv:1802.07228 [cs] [Preprint]. Available at: http://arxiv.org/abs/1802.07228 (Accessed: 23 November 2021).

Brundage, M. et al. (2020) ‘Toward Trustworthy AI Development: Mechanisms for Supporting Verifiable Claims’, arXiv:2004.07213 [cs] [Preprint]. Available at: http://arxiv.org/abs/2004.07213 (Accessed: 23 November 2021).

CDEI (2021) The roadmap to an effective AI assurance ecosystem. UK Government Centre for Data Ethics and Innovation. 

Cha, S. (2024) ‘Towards an international regulatory framework for AI safety: lessons from the IAEA’s nuclear safety regulations’, Humanities and Social Sciences Communications, 11(1), pp. 1–13. Available at: https://doi.org/10.1057/s41599-024-03017-1.

Cheng, D. (2024) Evaluating An AI Chip Registration Policy. Convergence Analysis. Available at: https://www.convergenceanalysis.org/research/evaluating-an-ai-chip-registration-policy.

Chowdhury, R. (2023) ‘AI Desperately Needs Global Oversight’, Wired, 6 April. Available at: https://www.wired.com/story/ai-desperately-needs-global-oversight/ (Accessed: 22 April 2024).

Cottier, B. (2023) Trends in the dollar training cost of machine learning systems. Epoch. Available at: https://epochai.org/blog/trends-in-the-dollar-training-cost-of-machine-learning-systems (Accessed: 15 March 2023).

Dohmen, H. and Feldgoise, J. (2023) ‘Explainer: The Commerce Department’s October 2023 Export Control Update -- A Bigger Yard, A Higher Fence: Understanding BIS’s Expanded Controls on Advanced Computing Exports’, Center for Security and Emerging Technology, 4 December. 

Dragan, A., King, H. and Dafoe, A. (2024) Frontier Safety Framework. Google DeepMind. Available at: https://deepmind.google/discover/blog/introducing-the-frontier-safety-framework/.

Drexel, B. and Depp, M. (2024) ‘Every Country Is on Its Own on AI’, Foreign Policy, 18 April. Available at: https://foreignpolicy.com/2023/06/13/ai-regulation-international-nuclear/ (Accessed: 17 April 2024).

DSIT (2023) Capabilities and risks from frontier AI: A discussion paper on the need for further research into AI risk. DSIT. 

DSIT (2024a) ‘Global leaders agree to launch first international network of AI Safety Institutes to boost cooperation of AI’, GOV.UK, 21 May. Available at: https://www.gov.uk/government/news/global-leaders-agree-to-launch-first-international-network-of-ai-safety-institutes-to-boost-understanding-of-ai (Accessed: 22 May 2024).

DSIT (2024b) ‘Seoul Declaration for safe, innovative and inclusive AI by participants attending the Leaders’ Session: AI Seoul Summit, 21 May 2024’, GOV.UK, 21 May. Available at: https://www.gov.uk/government/publications/seoul-declaration-for-safe-innovative-and-inclusive-ai-ai-seoul-summit-2024/seoul-declaration-for-safe-innovative-and-inclusive-ai-by-participants-attending-the-leaders-session-ai-seoul-summit-21-may-2024 (Accessed: 22 May 2024).

DSIT (2024c) ‘Seoul Ministerial Statement for advancing AI safety, innovation and inclusivity: AI Seoul Summit 2024’, GOV.UK, 22 May. Available at: https://www.gov.uk/government/publications/seoul-ministerial-statement-for-advancing-ai-safety-innovation-and-inclusivity-ai-seoul-summit-2024/seoul-ministerial-statement-for-advancing-ai-safety-innovation-and-inclusivity-ai-seoul-summit-2024 (Accessed: 22 May 2024).

DSIT (2024d) ‘UK \& United States announce partnership on science of AI safety’, GOV.UK, 2 April. Available at: https://www.gov.uk/government/news/uk-united-states-announce-partnership-on-science-of-ai-safety (Accessed: 18 April 2024).

Egeland, K. (2017) The road to prohibition: nuclear hierarchy and disarmament, 1968–2017. http://purl.org/dc/dcmitype/Text. University of Oxford. Available at: https://ora.ox.ac.uk/objects/uuid:b03d68ab-4748-4de7-a2e9-15616de6a05c (Accessed: 11 November 2021).

Fedasiuk, R., Elmgren, K. and Lu, E. (2022) Silicon Twist: Managing the Chinese Military’s Access to AI Chips. Center for Security and Emerging Technology. Available at: https://doi.org/10.51593/20210068.

Fischer, S.-C. and Wenger, A. (2019) ‘A Neutral Hub for AI Research’. Available at: https://doi.org/10.3929/ETHZ-B-000332541.

Fist, T. and Grunewald, E. (2023) Preventing AI Chip Smuggling to China. Washington DC: CNAS. Available at: https://www.cnas.org/publications/reports/preventing-ai-chip-smuggling-to-china (Accessed: 5 December 2023).

Gagnon, P. (2016) Who Cares about Particle Physics?: Making Sense of the Higgs Boson, the Large Hadron Collider and CERN. Oxford: OUP Oxford.

Gaithersburg Md Nist (2024) Managing Misuse Risk for Dual-Use Foundation Models. NIST AI NIST AI 800-1 ipd. Gaithersburg, MD: National Institute of Standards and Technology, p. NIST AI NIST AI 800-1 ipd. Available at: https://doi.org/10.6028/NIST.AI.800-1.ipd.

Grunewald, E. and Aird, M. (2023) AI chip smuggling into China: Potential paths, quantities, and countermeasures. Washington DC: IAPS.

Grunewald, E. and Fist, T. (2024) Comments on the Advanced Computing/Supercomputing IFR: Export Control Strategy \& Enforcement for AI Chips. CNAS. Available at: https://www.cnas.org/publications/commentary/comments-on-the-advanced-computing-supercomputing-ifr-export-control-strategy-enforcement-for-ai-chips.

Guterres, A. (2023) ‘Secretary-General Urges Security Council to Ensure Transparency, Accountability, Oversight, in First Debate on Artificial Intelligence | Meetings Coverage and Press Releases’, 18 July. 

Hausenloy, J. and Dennis, C. (2023) Towards a UN Role in Governing Foundation Artificial Intelligence Models. United Nations University.

Hausenloy, J., Miotti, A., and Dennis, C. (2023). Multinational AGI Consortium (MAGIC): A Proposal for International Coordination on AI.

Heim, L. et al. (2024) Governing through the cloud: the intermediary role of compute providers in AI regulation. Oxford Martin AI Governance Initiative. Available at: https://www.oxfordmartin.ox.ac.uk/publications/governing-through-the-cloud-the-intermediary-role-of-compute-providers-in-ai-regulation.

Heim, L. and Koessler, L. (2024) ‘Training Compute Thresholds: Features and Functions in AI Regulation’. arXiv. Available at: http://arxiv.org/abs/2405.10799 (Accessed: 8 August 2024).

Hickey, A. (2024) ‘The GPT Dilemma: Foundation Models and the Shadow of Dual-Use’. arXiv. Available at: https://doi.org/10.48550/arXiv.2407.20442.

Hirschman, A.O. (1970). Exit, Voice, and Loyalty: Responses to Decline in Firms, Organizations, and States. Harvard University Press.

Ho, L. et al. (2023) ‘International Institutions for Advanced AI’. arXiv. Available at: https://doi.org/10.48550/arXiv.2307.04699.

IAEA (2024) IAEA Open Knowledge Wiki - Safety Case, IAEA Open Knowledge Wiki - Safety Case. (Accessed: 8 August 2024).

Irving, G. (2024) ‘Safety cases at AISI’, AI Safety Institute, 23 August. Available at: https://www.aisi.gov.uk/work/safety-cases-at-aisi (Accessed: 13 September 2024).

Jewell, B. (2024) ‘Semiconductor CapEx Down in 2024’, Semiwiki, 3 August. Available at: https://semiwiki.com/semiconductor-services/343353-semiconductor-capex-down-in-2024/ (Accessed: 8 August 2024).

Kemp, L. et al. (2019) UN High-level Panel on Digital Cooperation: A Proposal for International AI Governance. Available at: https://www.cser.ac.uk/news/advice-un-high-level-panel-digital-cooperation/ (Accessed: 17 April 2024).

Kerry, C.F., Meltzer, J.P. and Renda, A. (2022) Global AI Cooperation on the Ground: AI R\&D on a global scale. Brookings.

Kremer, M., Levin, J. and Snyder, C.M. (2020) ‘Advance Market Commitments: Insights from Theory and Experience’, AEA Papers and Proceedings, 110, pp. 269–273. Available at: https://doi.org/10.1257/pandp.20201017.

La Commission de l’intelligence artificielle (2024) IA: notre ambition pour la France 25 propositions pour saisir les opportunités de l’intelligence artificielle. La Commission de l’intelligence artificielle. Available at: https://www.gouvernement.fr/actualite/25-recommandations-pour-lia-en-france (Accessed: 20 March 2024).

Law, H. and Ho, L. (2023) ‘Can a dual mandate be a model for the global governance of AI?’, Nature Reviews Physics, 5(12), pp. 706–707. Available at: https://doi.org/10.1038/s42254-023-00670-4.

LCFI (2023) ‘Do companies’ AI Safety Policies meet government best practice? - LCFI’, LCFI - Leverhulme Centre for the Future of Intelligence, 31 October. Available at: https://www.lcfi.ac.uk/news-events/news/ai-safety-policies (Accessed: 13 September 2024).

Linghan, Z. et al. (2024) ‘Artificial Intelligence Law of the People’s Republic of China (Draft for Suggestions from Scholars)’, Center for Security and Emerging Technology, 2 May. Available at: https://cset.georgetown.edu/publication/china-ai-law-draft/ (Accessed: 8 August 2024).

Maas, M.M. and Villalobos, J.J. (2023) ‘International AI Institutions: A Literature Review of Models, Examples, and Proposals’. Rochester, NY. Available at: https://doi.org/10.2139/ssrn.4579773.

Markhof, F.M. (2023) ‘IAEA Safeguards: A Model for International AI Regulation’, iGlobenews, 19 May. Available at: https://www.iglobenews.org/iaea-safeguards-a-model-for-international-ai-regulation/ (Accessed: 17 April 2024).

Mecklin, J. (2023) ‘Why the IAEA model may not be best for regulating artificial intelligence’, Bulletin of the Atomic Scientists, 9 June. Available at: https://thebulletin.org/2023/06/why-the-iaea-model-may-not-be-best-for-regulating-artificial-intelligence/ (Accessed: 17 April 2024).

METR (2023) ‘Update on ARC’s recent eval efforts: More information about ARC’s evaluations of GPT-4 and Claude’, 17 March. Available at: https://metr.org/blog/2023-03-18-update-on-recent-evals/ (Accessed: 3 September 2024).

Nevo, S. et al. (2024) Securing AI Model Weights: Preventing Theft and Misuse of Frontier Models. RAND Corporation.

O’Keefe, C. (2024) ‘Chips for Peace: How the U.S. and Its Allies Can Lead on Safe and Beneficial AI’, Default, 10 July. 

Patel, D. (2024) ‘Mark Zuckerberg - Llama 3, Open Sourcing \$10b Models, \& Caesar Augustus’, 28 March. Available at: https://www.dwarkeshpatel.com/p/mark-zuckerberg (Accessed: 22 April 2024).

Payne, K. (2023) ‘IAEA for AI? That won’t fly’, 18 August. Available at: https://www.kennethpayne.uk/p/iaea-for-ai-that-wont-fly (Accessed: 17 April 2024).

Pouget, H. et al. (2024) The Future of International Scientific Assessments of AI’s Risks. Carnegie Endowment for International Peace. Available at: https://carnegieendowment.org/research/2024/08/the-future-of-international-scientific-assessments-of-ais-risks?lang=en.

Pullella, P. (2023) ‘Pope Francis calls for binding global treaty to regulate AI’, Reuters, 14 December. Available at: https://www.reuters.com/technology/pope-calls-binding-global-treaty-artificial-intelligence-2023-12-14/ (Accessed: 12 August 2024).

Rhodes, R. (1986) The making of the atomic bomb. New York: Simon \& Schuster.

Roberts, H. et al. (2023) ‘Global AI governance: barriers and pathways forward’. Rochester, NY. Available at: https://doi.org/10.2139/ssrn.4588040.

Robinson, M. (2019) ‘The CERN Community; A Mechanism for Effective Global Collaboration?’, Global Policy, 10(1), pp. 41–51. Available at: https://doi.org/10.1111/1758-5899.12608.

Robinson, M. (2021) ‘Big Science Collaborations; Lessons for Global Governance and Leadership’, Global Policy, 12(1), pp. 66–80. 

Sandbrink, J.B. (2023) ‘Artificial intelligence and biological misuse: Differentiating risks of language models and biological design tools’. arXiv. Available at: https://doi.org/10.48550/arXiv.2306.13952.

Sastry, G. et al. (2024) ‘Computing Power and the Governance of Artificial Intelligence’. arXiv. Available at: https://doi.org/10.48550/arXiv.2402.08797.

Sevilla, J. et al. (2024) Can AI Scaling Continue Through 2030? Epoch AI. Available at: https://epochai.org/blog/can-ai-scaling-continue-through-2030 (Accessed: 21 August 2024).

Shavit, Y. (2023) ‘What does it take to catch a Chinchilla? Verifying Rules on Large-Scale Neural Network Training via Compute Monitoring’. arXiv. Available at: https://doi.org/10.48550/arXiv.2303.11341.

Sheehan, M. (2023) China’s AI Regulations and How They Get Made. Washington DC: Carnegie Endowment for International Peace.

Shevlane, T. et al. (2023) ‘Model evaluation for extreme risks’. arXiv. Available at: https://doi.org/10.48550/arXiv.2305.15324.

Stafford, E. and Trager, R.F. (2022) The IAEA Solution: Knowledge Sharing to Prevent Dangerous Technology Races. Centre for the Governance of AI (GovAI). Available at: https://www.governance.ai/research-paper/knowledge-sharing-to-prevent-dangerous-technology-races.

Stix, C. (2022) ‘Foundations for the future: institution building for the purpose of artificial intelligence governance’, AI and Ethics, 2(3), pp. 463–476. Available at: https://doi.org/10.1007/s43681-021-00093-w.

The Elders (2023) ‘The Elders urge global co-operation to manage risks and share benefits of AI’, 31 May. Available at: https://theelders.org/news/elders-urge-global-co-operation-manage-risks-and-share-benefits-ai (Accessed: 12 August 2024).

Trager, R. et al. (2023) ‘International Governance of Civilian AI: A Jurisdictional Certification Approach’. arXiv.

United States Department of Justice Office of Public Affairs (2023) Justice and Commerce Departments Announce Creation of Disruptive Technology Strike Force. United States Department of Justice.

Watson, M. (2023) ‘Opinion | IAEA for AI? That Model Has Already Failed’, Wall Street Journal, 1 June. Available at: https://www.wsj.com/articles/iaea-for-ai-that-model-has-already-failed-chaptgpt-technology-nuclear-proliferation-4339543b (Accessed: 17 April 2024).

Wold, J.W. (2024) ‘Von der Leyen gives nod to €100 billion “CERN for AI” proposal’, www.euractiv.com, 25 July. Available at: https://www.euractiv.com/
section/digital/news/von-der-leyen-gives-nod-to-e100-billion-cern-for-ai-proposal/ (Accessed: 8 August 2024).

Zhang, D. et al. (2022) Enhancing International Cooperation in AI Research: The Case for a Multilateral AI Research Institute. Stanford Institute for Human-Centered Artificial Intelligence. Available at: https://hai.stanford.edu/white-paper-enhancing-international-cooperation-ai-research-case-multilateral-ai-research-institute.

\end{document}